\documentclass[traditabstract]{aa}

\usepackage{graphicx}
\usepackage{txfonts}
\usepackage{multirow}
\usepackage{longtable}
\usepackage{natbib,twoopt}
\bibpunct{(}{)}{;}{a}{}{,} 

\begin{document}
\newcommand{\kms}{km\,s$^{-1}$}
%
%
  \title{Pure rotational spectra of TiO and TiO$_2$ \\in VY Canis Majoris\thanks{Based on observations carried out with the Submillimeter Array, IRAM Plateau de Bure Interferometer, and data acquired within the ESO program 67.B-0504.}}

  \author{
  T. Kami\'{n}ski\inst{\ref{inst1}}, C. A. Gottlieb\inst{\ref{inst2}}, 
  K. M. Menten\inst{\ref{inst1}}, N. A. Patel\inst{\ref{inst2}}, K. H. Young\inst{\ref
     {inst2}},\\ S. Br{\"u}nken\inst{\ref{inst3}}, H. S. P. M\"uller\inst{\ref{inst3}},
     M. C. McCarthy\inst{\ref{inst2}}, J. M. Winters\inst{\ref{inst4}}, L. Decin\inst{\ref{inst5},\ref{inst6}}    
  } 

  \offprints{T. Kami\'{n}ski}     

  \institute{\centering Max-Planck Institut f\"ur Radioastronomie, 
       Auf dem H\"ugel 69, 53121 Bonn, Germany,\\ \email{kaminski@mpifr.de}\label{inst1}
       \and Harvard-Smithsonian Center for Astrophysics, 60 Garden Street, Cambridge, MA, USA\label{inst2}
       \and I. Physikalisches Institut, Z\"ulpicher Strasse 77, 50937 K\"oln, Germany\label{inst3}
       \and Institut de Radioastronomie Millim{\'e}trique, 300 rue de la Piscine, 38406 Saint-Martin d'H{\'e}res, France\label{inst4}
       \and Instituut voor Sterrenkunde, Katholieke Universiteit Leuven, Celestijnenlaan 200D, 3001 Leuven, Belgium\label{inst5}
       \and Sterrenkundig Instituut Anton Pannekoek, University of Amsterdam, Science Park 904, 1098 Amsterdam, The Netherlands\label{inst6}
       }
	    
  \date{Received; accepted}

\abstract{We report the first detection of pure rotational transitions of TiO and TiO$_2$ at (sub-)millimeter wavelengths towards the red supergiant VY\,CMa. A rotational temperature, $T_{\rm rot}$, of about 250\,K was derived for TiO$_2$. Although $T_{\rm rot}$ was not well constrained for TiO, it is likely somewhat higher than that of TiO$_2$. The detection of the Ti oxides confirms that they are formed in the circumstellar envelopes of cool oxygen-rich stars and may be the ``seeds'' of  inorganic-dust formation, but alternative explanations for our observation of TiO and TiO$_2$ in the cooler regions of the envelope cannot be ruled out at this time. The observations suggest that a significant fraction of the oxides is not converted to dust, but instead remains in the gas phase throughout the outflow.} 

\authorrunning{Kami\'{n}ski et al.}
\titlerunning{Rotational spectra of TiO and TiO$_2$ in VY\,CMa}
\maketitle

\section{Introduction}\label{intro}

Copious amounts of dust, omnipresent in cool circumstellar environments, are produced by stars at the final stages of their 
evolution, yet  the formation of dust by depletion of molecules from the gas phase is poorly understood.  
In low-mass stars, the  formation of dust is thought to be most efficient in the asymptotic giant branch (AGB), and in massive stars 
during the red supergiant phase.
There are two major types of dust produced by cool stars. Carbon stars (e.g., C-type AGB stars and R\,CrB stars), which have more carbon than oxygen in their circumstellar environments, produce mainly carbon dust which nucleates from abundant aromatic hydrocarbons \citep[][and references therein]{carbondust}. In oxygen-rich circumstellar envelopes (e.g., M-type AGB stars and red supergiants), essentially all carbon is locked into CO and the dust is inorganic. How such inorganic grains start to form remains elusive.  
After ruling out species containing the more abundant elements (Si, Fe, and Mg) and taking into account infrared (IR) observations of circumstellar dust shells in cool oxygen-rich giants and supergiants, \citet{inorganicdust} concluded on the basis of a chemical equilibrium model that the formation of inorganic dust at higher temperatures entails molecules containing less abundant elements.
Gail \& Sedlmayr showed that at temperatures expected at the onset of dust formation, $\sim$1200\,K, the refractory molecules TiO and TiO$_2$ are the first species in the chain of reactions that produce titanium oxide clusters which are the ``seeds'' needed to initiate further dust formation at temperatures below the condensation temperature of TiO \citep[see also][]{jeong}. 

Prior to this work, TiO$_2$ had not been identified in an astrophysical object either at optical or radio wavelengths, but TiO has been observed at visible wavelengths in M-type stars for many decades \citep[][and references therein]{merrill}. The electronic bands of TiO are so prominent in late-type stars, that the Morgan-Keenan spectroscopic classification is based in part on the relative strengths of these bands \citep{MK}. However, the electronic bands in these studies are {\it photospheric} absorption features, which do not provide any insight into the warm circumstellar chemistry. Occasionally the optical bands of TiO are observed in emission from the warm circumstellar gas, but such cases are extremely rare and difficult to interpret \citep{uequ,kami_v4332}. Perhaps the most extreme example of the importance of TiO as an opacity source in oxygen-rich stars at optical wavelengths are high-amplitude mira variables, for which molecular absorption of TiO may be responsible for the variation of optical light by up to 8 mag \citep{reidGoldstone}.

Radio astronomers had long sought TiO in the interstellar medium and circumstellar environments \citep{churchwell,millar87,TiO2freq1}, because observations of its rotational lines might yield the temperature and abundances more directly than at optical wavelengths. Although there have been several attempts to observe TiO at radio wavelengths (most with single antennas), all were unsuccessful.  In the past few years it was realized that images obtained with sensitive interferometers with sub-arcsecond resolution might have sufficient sensitivity to observe  the pure rotational spectrum of TiO and possibly TiO$_2$. 

The rotational spectrum of TiO was measured in the laboratory many years ago \citep{TiOfreq}, and the essential laboratory spectroscopic measurements needed  to identify TiO$_2$ at optical and radio wavelengths were reported more recently.  The rotational spectrum of TiO$_2$ was first observed in the laboratory when \cite{TiO2freq1} measured 13 transitions to high accuracy in the centimeter band.  Br{\"u}nken et al. determined the rotational and leading centrifugal distortion constants that allow predictions of the astronomically important transitions in the millimeter and submillimeter bands which subsequent measurements confirmed are accurate to 0.5\,\kms\ or better \citep{TiO2freq2}.
Unfortunately,  the prospect for identifying TiO$_2$ at optical wavelengths appears difficult, because the most promising band at 
$\sim$5300\,\AA\ may be obscured by a more intense band of TiO.

The red supergiant  VY\,CMa appeared to be the best source to search for TiO and TiO$_2$ at 
 (sub-)millimeter wavelengths, because  optical emission bands of refractory oxides including ScO, YO, VO, and TiO have long been known to be present in this extreme star;\footnote{Approximate luminosity  \citep[scaled to a distance of 1.2\,kpc,][]{bo} is 3$\cdot$10$^5$\,L$_{\sun}$, effective temperature $T_{\rm eff}$=3200--3500\,K, and radius $R_{\rm \star}$=1500--3000\,R$_{\rm \sun}$ (see Appendix in \citealp{humphreys07}; \citealp{wittkowski}).} and the (sub-)millimeter spectrum of VY\,CMa is unusually rich for an oxygen-rich source.
In recent spectral line surveys at millimeter wavelengths at the Arizona Radio Observatory \citep[e.g.,][]{ziurys,tanen10}, the number of detected molecules in VY\,CMa increased to 21 including metal oxides (e.g., AlO and PO) and other exotic molecules (e.g., AlOH and PN). Some of the newly discovered molecules are refractory species important for dust formation in oxygen-rich CSEs.

In 2010, we made a wide band line survey of VY\,CMa in the 345-GHz band with the Submillimeter Array (SMA). Present in this survey are rotational lines of both TiO and TiO$_2$. The identifications  of TiO and TiO$_2$ were confirmed in 2012 by independent measurements with the IRAM Plateau de Bure Interferometer (PdBI) of lines within the 220.7--224.3\,GHz range. In Sect.\,\ref{obs}, we briefly describe the SMA and PdBI observations; in Sect.\,\ref{ana}, the line identification and properties of the TiO and TiO$_2$ emission are presented; in Sect.\,\ref{rot-diagrams}, the rotational-temperature diagrams  for both molecules are analyzed; in Sect.\,\ref{optical}, the (sub-)millimter TiO emission is compared with optical observations of this molecule; and finally, in the Sect.\,\ref{dis}, the observed properties of the emission are discussed within the broad context of inorganic dust formation. 


\section{Observations}\label{obs}

Rotational lines of the two simplest Ti oxides (TiO and TiO$_2$) were detected in a 345\,GHz-band spectral line survey with the SMA\footnote{The raw survey data are available at http://www.cfa.harvard.edu/cgi-bin/sma/smaarch.pl}. The results of the survey together with a detailed description of the observations and data reduction will be described elsewhere.  Here we provide a brief description of the observations.

The spectral line survey of VY\,CMa covers the entire frequency range between 279.1 and 355.1\,GHz, except for very narrow 32\,MHz wide gaps at every 2\,GHz interval. The observations were done between 17 January and 2 February 2010. The typical time spent in each 4\,GHz wide band was about 3.9\,h, and the rms noise for spectra extracted with an aperture of 1\arcsec\ was about 70 mJy per resolution element of 0.7\,\kms. The noise is fairly uniform over the entire range of the survey except for the region between 321--327\,GHz, owing to a strong telluric absorption band of water. 

The SMA was used in its extended configuration to produce a synthesized beam with a full width at half-maximum (FWHM) of about $0\farcs$9. The phase center was at $\alpha(2000)$=$07^{\rm{h}}22^{\rm{m}}58.332^{\rm{s}},
\delta(2000)$=$-25^{\circ}46^{\prime}03.17\farcs2$. All necessary calibrations (gain, flux, bandpass) were secured during the observations. The data were calibrated in the MIR-IDL package\footnote{http://cfa-www.harvard.edu/$\sim$cgi/mircook.html}. After a careful inspection of the flux calibration, we estimate that its typical uncertainty is about 15\%. The continuum map of VY\,CMa (0.6\,Jy) was used to self-calibrate the data in phase and to correct for some of the phase variations. Most of the further data processing, including map production from calibrated visibilities, was performed in Miriad \citep{miriad}. 


To confirm our identification of TiO and TiO$_2$ with the SMA, we made an independent observation of a 3.6\,GHz-wide band near 222.45\,GHz with the PdBI on 22 April and 13 October 2012.  The frequency range (220.65--224.25\,GHz) was chosen so as to maximize the number of unblended and relatively strong lines of TiO and TiO$_2$ which could be observed simultaneously in the 1.35\,mm band. The PdBI was used in the D configuration with six antennas. Owing to the low declination of VY\,CMa, the maximum elevation is only 19\degr.  As a result, the observations were affected by antenna shadowing and the synthesized beam is very asymmetric (4\farcs9$\times$2\farcs2, with a position angle of the major axis of 158\degr). The observations of VY\,CMa were interleaved every 22\,min by short integrations on nearby phase and amplitude calibrators (first 0648-165 and 0646-306, then switching to 0648-165 and 0823-223 when the elevation of 0646-306 was too low). Pointing and focus were checked about every 65\,min. The observations were made with the dual-polarization single-sideband receiver (tuned in the lower side band) and the WideX correlator, which provides a resolution of 2\,MHz (corresponding to 2.7\,\kms\ at the observing frequency). The data were calibrated using the GILDAS\footnote{http://www.iram.fr/IRAMFR/GILDAS} software package and the final processing included self-calibration in phase on the strong continuum of VY\,CMa. The thermal noise in the combined data set is $\sim$8 mJy/beam per 2.7\,\kms\ bin.


\section{Line identification and measurements}\label{ana}

In all, 220 spectral lines were observed in the SMA survey. Initially we were able to assign all except about 30 on the basis of the close coincidence between the astronomical frequency and laboratory rest frequencies ($\nu_{\rm rest}$) tabulated in the standard catalogs (Cologne Database for Molecular Spectroscopy, CDMS, \citealp{CDMS1,CDMS2}; Jet Propulsion Laboratory Catalogue, \citealp{JPL}). Of the remaining unassigned features, lines of TiO were relatively easy to recognize in the spectrum because they were narrower than most others, while TiO$_2$ was one of the last species considered in our iterative identification procedure. 

The full spectrum obtained with PdBI is shown and described in Appendix\,\ref{ap1}. All of the lines of TiO and TiO$_2$ predicted to be present in the spectrum are detected in these very sensitive observations.


Below, we describe the identification and present the spectra of TiO and TiO$_2$. Data from the SMA survey are presented first and are followed by a description of the PdBI data, as this is the chronological order that influenced the analysis. A search for isotopologues in all the available data is described in Sect.\,\ref{iso}. Some constraints on the emission angular sizes and position are given in Sect.\,\ref{maps}. The evidence in support of the identifications is summarized in Sect\,\ref{confi}.

\subsection{TiO in the SMA survey}\label{ana1}

The TiO radical has a X$^3\Delta$ electronic ground state that is split into three fine structure components with energies relative to the lowest $^3\Delta_1$ component of 140\,K ($^3\Delta_2$) and 280\,K ($^3\Delta_3$). The rotational spectrum of TiO had been measured to high accuracy in the laboratory by \citet{TiOfreq} and the dipole moment was measured recently \citep[$\mu$=3.34\,D;][]{steimle}.
Starting with the $J$=3$\to$2 transition, roughly every 32\,GHz there is a rotational transition ($J^{\prime}\!\!\to\!\!J$) separated by 3--4\,GHz from the corresponding transition in the next lower or higher fine structure component. Eight  transitions of TiO were covered in the survey: all three components in $J$=9$\to$8 and 10$\to$9, but only two components were covered in $J$=11$\to$10 because the third (at 355.6\,GHz) lies slightly outside the upper end of the observed range.

Six lines of TiO were detected above the 3$\sigma$ level in the SMA survey. Maps of the integrated emission, confirm that all six lines are from a source centered (within uncertainties) on the star (see Sect.\,\ref{maps}). Of the two missing transitions, the line at 291.0\,GHz is located close to a gap and the baseline in the region near the line is uncertain. If the baseline in this spectrum is removed in a least squares fit with a polynomial, the edge of the spectrum near the gap is slightly lower and the fluxes in the frequency range of the expected feature of TiO are greater than 3 times the local rms noise.
However, in the following discussion we treat the line at 291.0~GHz as though it had not been detected, and the measurements cited here are for the spectrum with a linear baseline subtracted (see below). The line at 323.3\,GHz ($J$=10$\to$9 in $^3\Delta_3$) is located in the frequency region affected by the absorption band of terrestrial water and the noise is much greater than the expected intensity. As a result, none of the lines of the $^3\Delta_3$ component was definitely detected.

The most intense line of TiO at 348.2\,GHz may be blended with very weak emission from a line of $^{34}$SO$_2$ at 348.117\,GHz. Approximately 30 lines of SO$_2$ were observed in the survey, and  from these a rotational temperature of 85$\pm$10\,K was derived.  On the assumption that $T_{\rm{rot}}$ is the same for $^{34}$SO$_2$, we estimate that the contribution from the possibly interfering line of $^{34}$SO$_2$ is $<$5\% of the measured fluxes of TiO. We cannot rule out {\it a priori} that  the lines of TiO  (and those of TiO$_2$, see Sect.\,\ref{ana2}) might be blended with species whose carriers have not been identified in the survey, nor can we exclude entirely the possibility of blends with weak lines from species that have been identified.  However the contribution to the measured fluxes from the latter should not be significant, because of the small aperture used in our analysis (1\arcsec$\times$1\arcsec). For example, in a molecule like SO$_2$, which has intense broad lines throughout the entire spectrum, the emission is greatest for larger apertures and transitions between the lowest rotational energy levels \citep[cf.][]{fu}. 

\begin{figure}\centering
\includegraphics[angle=0,width=\columnwidth]{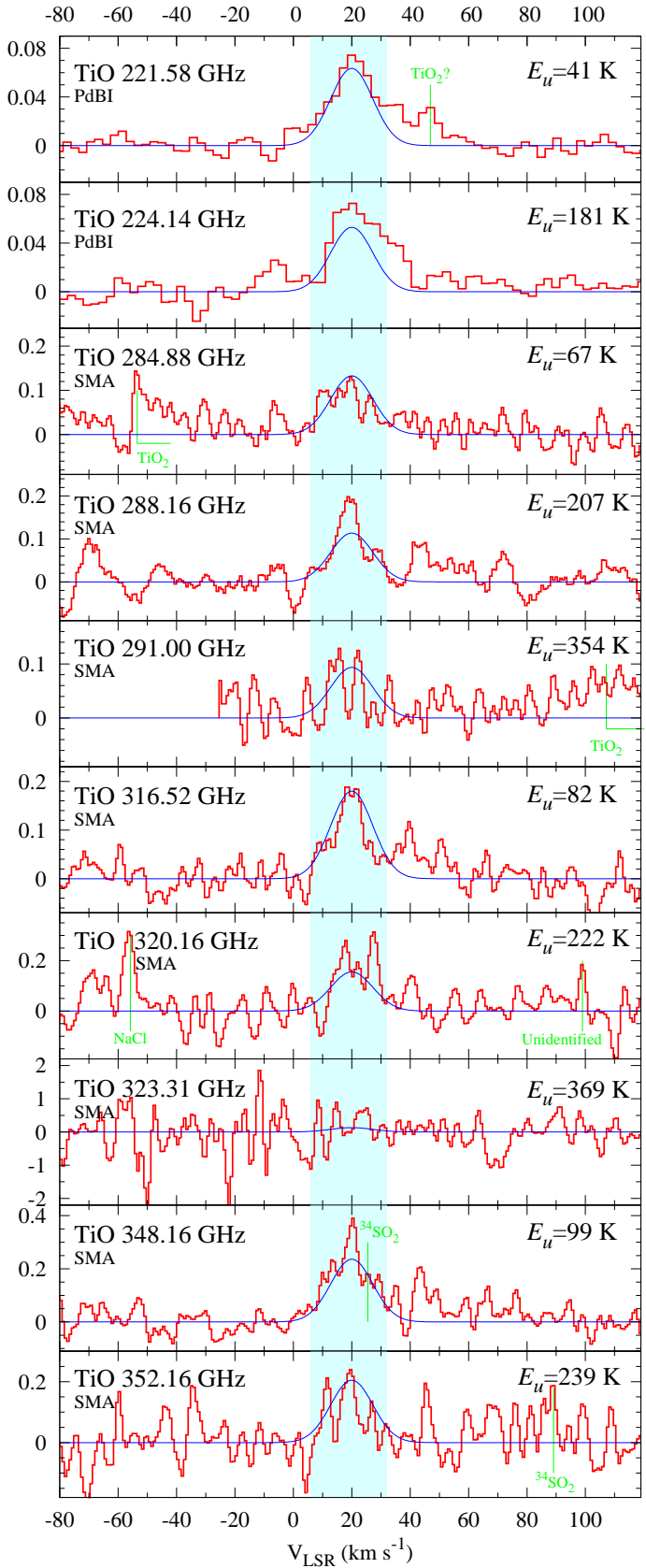}
\caption{Spectra of all transitions of TiO covered in the SMA survey and the PdBI spectrum. The shaded area is the range over which the spectra were integrated. Overlayed on the measured spectra are theoretical Gaussian profiles (in blue) plotted on the assumption of LTE at 1000\,K. The ordinate is flux density in Jy. Spectra correspond to the aperture of 1\arcsec$\times$1\arcsec\  (SMA data) or to the full PdBI beam of 4\farcs9$\times$2\farcs2 (PdBI data). The SMA spectra were Hanning smoothed.}
\label{Fig-indiv-TiO}
\end{figure}

\begin{figure}\centering
\includegraphics[angle=0,width=\columnwidth]{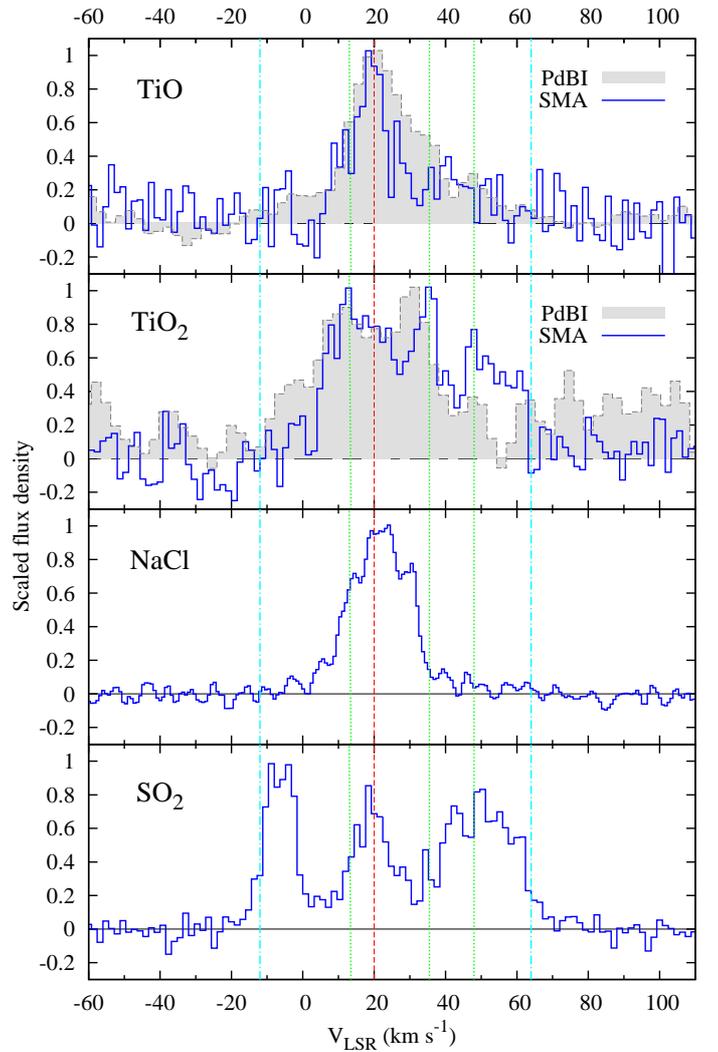}
\caption{Two top panels: Spectra of TiO and TiO$_2$ obtained by combining several lines from the SMA survey (blue line) or PdBI observations (filled grey histogram). They are compared to a combined profile of NaCl emission (third panel) and a strong line of SO$_2$ (16$_{0,16}$\,$\to$15$_{1,15}$) from the SMA survey (bottom panel). Spectra were smoothed and the peak flux density of each was normalized to 1.0. The dashed (red) vertical line marks the line center of TiO at $V_{\rm {LSR}}$=20\,\kms, and the dotted (green) vertical lines mark 14, 35.5, and 48.0\,\kms\ identified as peaks of velocity components in the SMA combined profile of TiO$_2$. The outermost dash-dotted lines (cyan) show the full velocity range of the TiO$_2$ emission.}
\label{Fig-averprofile}
\end{figure}

Listed in Table\,\ref{Tab-TiO} are the transitions of TiO which were covered with SMA and PdBI, their peak and integrated fluxes, the rms of the noise in the region near the lines, and the detection levels (signal-to-noise ratios, S/N). The latter are expressed as the integrated flux density, $I$, divided by the uncertainty, $\Delta I$, resulting from the 1\,rms noise level, i.e., $\Delta I$=rms$\cdot \Delta V_{\rm ch}\!\cdot\!\sqrt{n_{\rm ch}}$,  where $\Delta V_{\rm ch}$ is the channel width for which the rms is specified and $n_{\rm ch}$ is the number of channels within the integrated interval. For those transitions that were not detected, the 3$\sigma$ upper limit on the fluxes is indicated. Spectra of the individual lines are shown in Fig.\,\ref{Fig-indiv-TiO}, while Fig.\,\ref{Fig-averprofile} shows a combined profile of the lines that were detected with SMA, excluding the blended line at 348.16\,GHz. The combined spectrum is an average of individual spectra weighted by the rms$^{-2}$ and scaled by the reciprocal of the integrated flux. 

\begin{table} 
\caption{Observed parameters of the TiO lines. }\label{Tab-TiO}
\begin{tabular}{@{}c c r c@{~~} c@{}c@{ }c@{}r@{}}
\hline
Transition\tablefootmark{a}&
$\nu_{\rm rest}$&
\multicolumn{1}{c}{$E_{u}$}& 
$S_{ul}$\tablefootmark{b}&
Peak\tablefootmark{c}&
Flux\tablefootmark{d}&
 rms\tablefootmark{e}&
\multicolumn{1}{c@{}}{S/N\tablefootmark{f}}\\
$J^{\prime}\!\to\!J$, $\Omega$
       &(GHz)    &\multicolumn{1}{c}{(K)}&&(mJy)&(Jy\,\kms)&(mJy) &   \\           
\hline\hline 
\multicolumn{8}{c}{PdBI}\\
\hline
~~7$\to$6, ~~1~&   221.580 &  41&13.71 & 74 &~~~1.2    &  5.4 &  26.3\\
~~7$\to$6, ~~2~&   224.139 & 181&12.85 & 73 &~~~1.3    &  7.3 &  20.5\\
\hline
\multicolumn{8}{c}{SMA}\\
\hline
~~9$\to$8, ~~1~&   284.875 &  67&17.77 & ~~150 &~~~1.8    &  46 &  8.5\\
~~9$\to$8, ~~2~&   288.156 & 207&17.11 & ~~240 &~~~2.3    &  66 &  7.6\\
~~9$\to$8, ~~3~&   290.998 & 354&16.00 &$<$160 &~$<$0.7~  &  50 &  $<$3.0\\
10$\to$9, ~~1~ &   316.519 &  82&19.80 & ~~230 &~~~2.4    &  50 & 10.6\\
10$\to$9, ~~2~ &   320.159 & 222&19.20 & ~~580 &~~~3.1    & 174 &  3.9\\
10$\to$9, ~~3~ &   323.314 & 369&18.20 &$<$1470 &$<$14.0~& 1021 &$<$3.0\\
11$\to$10, 1   &   348.160 &  99&21.81 & ~~440 &~~~4.7    &  73 & 14.1\\
11$\to$10, 2   &   352.158 & 239&21.27 & ~~290 &~~~2.3    & 125 &  4.1\\
\hline\end{tabular}\tablefoot{
\tablefoottext{a}{$\Omega = 1, 2, 3$ refers to the fine structure components $^3\Delta_1, ^3\Delta_2$,  or $^3\Delta_3$.}
\tablefoottext{b}{The quantum mechanical line strength of the rotational transition.} 
\tablefoottext{c}{Peak flux density at a resolution of 2.7\,\kms\ (PdBI) or 0.8\,\kms\ (SMA) for the full synthesized beam.}
\tablefoottext{d}{Flux integrated between 6 and 32\,\kms\ and within the full synthesized beam.}
\tablefoottext{e}{The rms per bin of 2.7\,\kms\ (PdBI) or 0.8\,\kms\ (SMA).}
\tablefoottext{f}{The signal-to-noise ratio.}} 
\end{table}

The line profile of TiO as seen in the SMA combined spectrum alone, extends over the range from about 3 to $\ge$55\,\kms\ (Fig.\,\ref{Fig-averprofile}). Most of the emission occurs  between $\sim$3 and $\sim$32\,\kms\  and is well reproduced by a Gaussian profile centered at $V_{\rm LSR}$=20$\pm$2\,\kms\ and with a FWHM=18$\pm$2\,\kms. 
In addition, there is emission that forms a wing that is extended to the red of the main component. The redshifted emission 
is not discernible in the spectra of individual lines of TiO observed with SMA (Fig.\,\ref{Fig-indiv-TiO}).


\subsection{TiO in the PdBI data}\label{ana_pdb_tio}
%
Two lines of TiO were covered by the PdBI observation are both well detected: the $J$=7$\to$6 transition  in the $^3\Delta_1$ and $^3\Delta_2$ spin components at 221.6\,GHz 224.1\,GHz. The red wing of the 221.6\,GHz line may be very weakly contaminated by TiO$_2$ emission (see below) and the blue wing of the 224.1\,GHz line may be affected by vibrationally excited SO$_2$ (unlikely; see Appendix\,\ref{ap1}). The spectra of all TiO lines are shown in Fig.\,\ref{Fig-indiv-TiO} and their parameters are listed in Table\,\ref{Tab-TiO}. The line at 221.6\,GHz appears more triangular in shape than that at 224.1\,GHz, possibly owing to the much different excitation energies of 41 and 181\,K. 

The average line profile of the PdBI lines is compared to the combined SMA profile in Fig.\,\ref{Fig-averprofile}. The PdBI profile is slightly broader than that observed with SMA. In the more sensitive PdBI observations, not only the existence of the red wing is confirmed but also a blue wing is apparent. They form a broad ``pedestal'' extending from --15\,\kms\ to 65\,\kms\ (with an uncertainty of about 5\,\kms\ in each of the limits). Because the ratio of the central, Gaussian-like component to the broad pedestal is smaller in the PdBI data, the main component appears broader. 

\subsection{TiO$_2$ in the SMA data}\label{ana2}

Titanium dioxide, TiO$_2$, is an asymmetric top with $C_{\rm 2v}$ symmetry and a substantial dipole moment along the $b$ inertial axis 
\citep[$\mu_b$=6.33\,D;][]{wang}. Owing to the two equivalent off-axis $^{16}$O bosons, half of the rotational levels are missing. Because it is a fairly heavy molecule, many transitions lie within the range of the survey. 
We found 20 unblended lines and three pairs of lines that form tight blends of TiO$_2$ emission. The blend at 337.2\,GHz is the most intense feature of TiO$_2$ in the survey and one that provided the first indication that we may have detected this elusive molecule. In the blends at 331.2 and 337.2\,GHz, the flux from each line in the blended pair is nearly the same, because both arise from levels at nearly the same energy above ground and the line strengths are comparable. 

\begin{figure*}\centering
\includegraphics[angle=270,width=\textwidth]{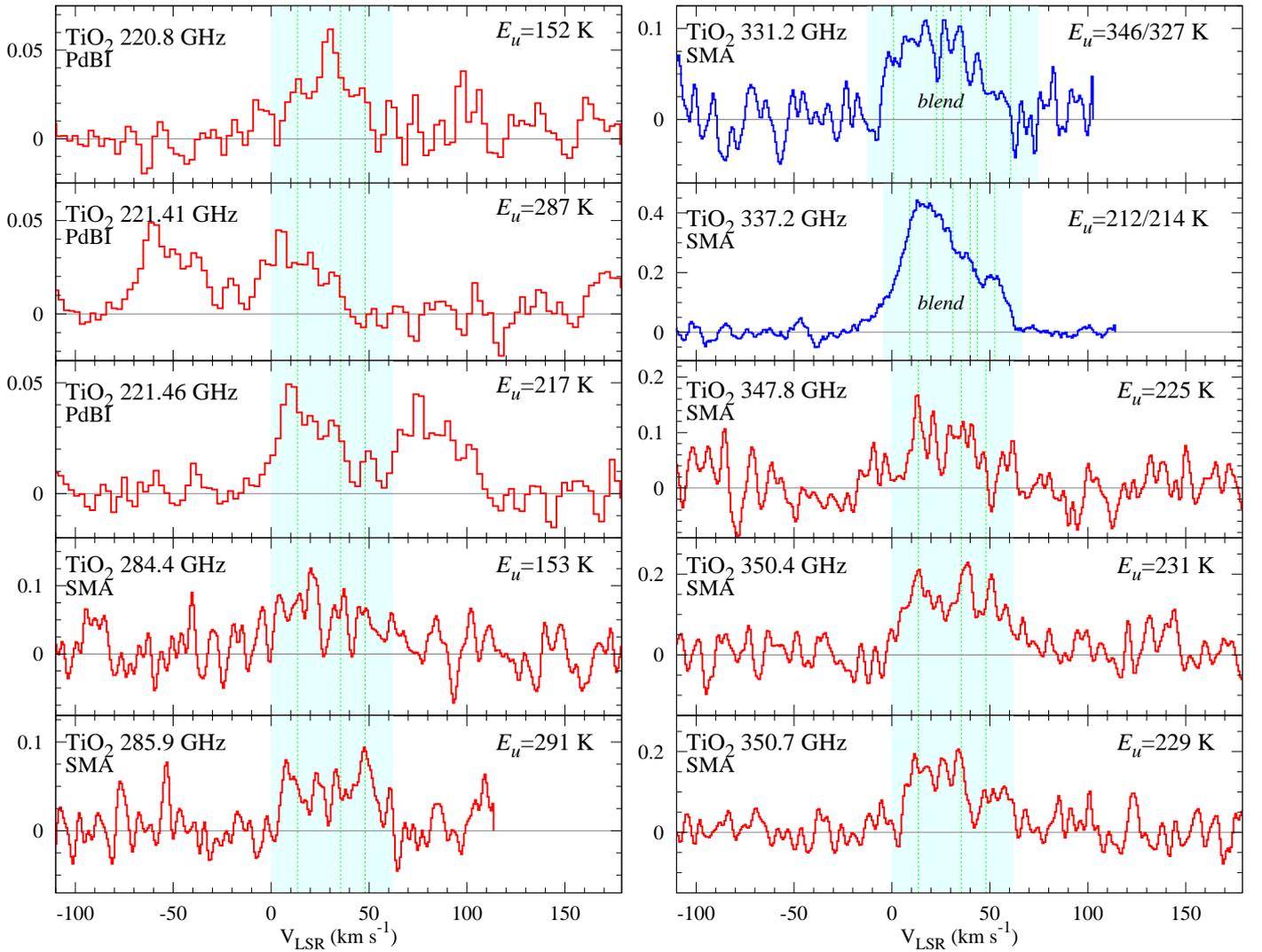}
\caption{Spectra of unblended lines of TiO$_2$ with the the highest S/N, and two blended lines (blue). The velocity scales of the blended lines refer to the average rest frequency of the two components. Parts of the spectra that contain strong emission from other species were blanked for clarity. The lines found in the PdBI observations are shown in the top three panels on the left. The dotted (green) vertical lines mark the position of the three main components of the TiO$_2$ emission identified in the combined profile (cf. Fig.\,\ref{Fig-averprofile}). The ordinate is flux density in Jy and corresponds to the aperture of 1\arcsec$\times$1\arcsec\ in the SMA spectra or to the full beam size (4\farcs9$\times$2\farcs2) in the PdBI data. The SMA spectra were Hanning smoothed.}
\label{Fig-indiv-TiO2}
\end{figure*}

Table\,\ref{Tab-TiO2} lists the transitions detected, their flux, and the rms noise. It also contains the uncertainties of the rest frequencies, which typically are below 0.1\,MHz, but in a few cases are larger than 1\,MHz. Shown in Fig.\,\ref{Fig-indiv-TiO2} are the features with the highest S/N, including the blended features at 331.2 and 337.2\,GHz. Four unblended lines with the highest S/N (at 285.9, 347.8, 350.4, and 350.7\,GHz) were combined to yield an average profile by the same procedure as that for TiO (Fig.\,\ref{Fig-averprofile}). The line profile of TiO$_2$ is broad, with most features extending over the velocity range from about 0 to 62\,\kms. The strong blend at 337.2\,GHz suggests, however, that the range may extend as far as --15\,\kms. {\it At least} three velocity components  at about 14, 35.5, and 48.0\,\kms\ are apparent in the combined spectrum.

\begin{table*} \centering 
\caption{Parameters of detected TiO$_2$ lines.   }\label{Tab-TiO2}
\begin{tabular}{ccrcccr}
\hline
Transition&$\nu_{\rm rest}$\tablefootmark{a}&
\multicolumn{1}{c}{$E_{u}$}& 
$S_{ul}$&
Flux\tablefootmark{b}&
 rms\tablefootmark{c}&
\multicolumn{1}{c@{}}{S/N\tablefootmark{d}}\\
$J{^{\prime}}_{K_a^{\prime},K_c^{\prime}}\!\to\!J_{K_a,K_c}$
       &(MHz)    &\multicolumn{1}{c}{(K)}&&(Jy\,\kms)&(mJy) & \\           
\hline\hline
\multicolumn{6}{c}{PdBI}  \\
\hline
23$_{5,19}\!\to$23$_{4,20}$  & 220812.1 (0.2) & 230 & 12.857 & 1.48 &  11.0 & 10.4\\
 4$_{4,0}\!\to$3$_{3,1}$     & 221095.7 (0.0) &  25 &~~3.476&$<$1.77~~~& ~~8.3 & 16.5\\
32$_{5,27}\!\to$32$_{4,28}$  & 221406.6 (0.5) & 424 & 19.365 & 0.91 & ~~7.5 &~~9.4\\
28$_{4,24}\!\to$28$_{3,25}$  & 221458.4 (0.3) & 323 & 14.873 & 1.45 & ~~7.5 & 14.9\\

\hline
\multicolumn{6}{c}{SMA}  \\
\hline
10$_{3,7}\!\to$ 9$_{2,8}$   &282196.8 (0.0) &  50  &~~2.489 &  ~~1.08 & 46.1 & 3.3  \\
21$_{1,21}\!\to$20$_{0,20}$ &284371.9 (0.1) &  153 & 19.115 & ~~3.51 & 49.5 & 10.1 \\
20$_{2,18}\!\to$19$_{3,17}$ &284951.3 (0.0) &  161 &~~9.737 &  ~~2.25 & 47.2 & 6.8  \\
25$_{7,19}\!\to$25$_{6,20}$ &285859.7 (0.4) &  291 & 14.106 & ~~2.73 & 42.1 & 9.2  \\
22$_{7,15}\!\to$22$_{6,16}$ &287253.6 (0.2) &  239 & 11.902 &  ~~2.41 & 55.5 & 6.2  \\
21$_{7,15}\!\to$21$_{6,16}$ &290830.5 (0.2) &  222 & 11.157 & \multirow{2}*{$\Big\}$~2.39\tablefootmark{e}}& \multirow{2}*{52.5} & \multirow{2}*{4.7}\\
42$_{9,33}\!\to$42$_{8,34}$ &290858.8 (7.8) &  753 & 28.013 &  &  & \\ 
19$_{7,13}\!\to$19$_{6,14}$ &293001.6 (0.1) &  192 &~~9.733 &  ~~2.69 & 59.5 & 6.4  \\
33$_{6,28}\!\to$33$_{5,29}$ &294443.0 (0.0) &  452 & 17.672 &  ~~1.68 & 59.5 & 4.0  \\
16$_{7,9}\!\to$16$_{6,10}$  &295303.4 (0.0) &  153 &~~7.651 &  ~~1.59 & 46.7 & 4.8  \\
19$_{3,17}\!\to$18$_{2,16}$ &304189.9 (0.0) &  147 &~~8.886 &  ~~2.08 & 64.1 & 4.6  \\
22$_{1,21}\!\to$21$_{2,20}$ &310554.7 (0.0) &  180 & 16.308 &  ~~3.39 & 82.4 & 5.8  \\
23$_{1,23}\!\to$22$_{0,22}$ &310782.7 (0.0) &  182 & 21.124 &  ~~5.03 & 84.3 & 8.5  \\
27$_{8,20}\!\to$27$_{7,21}$ &331211.8 (0.7) &  346 & 14.862 & \multirow{2}*{$\Big\}$~3.71}& \multirow{2}*{48.9} & \multirow{2}*{8.1}\\
26$_{8,18}\!\to$26$_{7,19}$ &331240.1 (0.6) &  327 & 14.125 &  &  & \\ 
39$_{8,32}\!\to$39$_{7,33}$ &331381.8 (2.5) &  644 & 23.430 &  ~~1.33 & 48.3 & 3.9  \\
12$_{3,9}\!\to$11$_{2,10}$  &331599.6 (0.0) &  68  &~~1.991 &  ~~1.14 & 52.6 & 3.1  \\
24$_{3,21}\!\to$23$_{4,20}$ &336469.4 (0.1) &  235 &~~9.160 &  ~~3.77 & 65.4 & 8.2  \\
25$_{1,25}\!\to$24$_{0,24}$ &337196.1 (0.1) &  214 & 23.128 & \multirow{2}*{$\Big\}$~16.97}& \multirow{2}*{54.2} & \multirow{2}*{37.2}\\
24$_{1,23}\!\to$23$_{2,22}$ &337206.1 (0.0) &  212 & 18.345 &  &  & \\ 
16$_{8,8}\!\to$16$_{7,9}$   &341875.8 (0.1) &  169 &~~7.078 &  ~~1.80 & 81.1 & 3.1  \\
12$_{4,8}\!\to$11$_{3,9}$   &345580.9 (0.0) &  75  &~~3.995 &  ~~1.43 & 53.7 & 3.8  \\
24$_{2,22}\!\to$23$_{3,21}$ &347788.1 (0.1) &  225 & 14.197 &  ~~4.08 & 69.0 & 8.4  \\
35$_{5,31}\!\to$35$_{4,32}$ &348661.4 (1.3) &  487 & 14.380 &  ~~1.68 & 65.6 & 3.6  \\
26$_{0,26}\!\to$25$_{1,25}$ &350399.0 (0.1) &  231 & 24.130 &  ~~8.19 & 71.4 & 16.3 \\
25$_{2,24}\!\to$24$_{1,23}$ &350707.9 (3.7) &  229 & 19.361 &  ~~6.73 & 70.2 & 13.6 \\
\hline
\multicolumn{6}{c}{ SMA deblended\tablefootmark{f} }  \\
\hline
27$_{8,20}\!\to$27$_{7,21}$ &331212.8 (0.7) & 346 & 14.862 & 1.87 & 48.9 &  4.6 \\ 
26$_{8,18}\!\to$26$_{7,19}$ &331240.1 (0.6) & 327 & 14.125 & 1.84 & 48.9 &  4.5  \\
25$_{1,25}\!\to$24$_{0,24}$ &337196.1 (0.1) & 214 & 23.128 & 9.29 & 54.2 &  22.8 \\
24$_{1,23}\!\to$23$_{2,22}$ &337206.1 (0.0) & 212 & 18.345 & 7.67 & 54.2 &  18.8 \\
\hline\end{tabular}\tablefoot{
\tablefoottext{a}{Rest frequency (and its error).}
\tablefoottext{b}{Flux integrated between 0 and 62\,\kms\ within the full synthesized beam.}
\tablefoottext{c}{The rms per bin of 2.7\,\kms\ (PdBI) or 0.8\,\kms\ (SMA).}
\tablefoottext{d}{The signal-to-noise ratio.} 
\tablefoottext{e}{Lines are not included in Fig.\,\ref{Fig-rotdiagr-TiO2}.}   
\tablefoottext{f}{Parameters for the deblended lines (see Sect.\,\ref{ana2}). }
}
\end{table*}
\subsection{TiO$_2$ in the PdBI data}\label{ana_pdb_tio2}

The Cologne Database for Molecular Spectroscopy lists eight lines of TiO$_2$ in the frequency range covered by the PdBI observations. They are marked in Fig.\,\ref{Fig-PdBI}, where the length of the markers indicates the relative line intensities expected under local thermodynamic equilibrium (LTE) conditions at 300\,K,  which is approximately the gas rotational temperature derived from the survey data (Sect.\,\ref{rot-diagrams}). Four are present in the observed spectrum. The one at 221.1\,GHz  is blended with emission of $^{34}$SO$_2$, which we estimate is weaker than the $^{34}$SO$_2$ line at 221.7\,GHz; therefore, the feature at 221.1\,GHz is dominated by TiO$_2$ emission. Two other lines, at 221.41 and 221.45\,GHz (i.e., 70\,\kms\ apart) overlap slightly; and another two, at 220.9 and 223.5\,GHz, are not detected, but are localized in spectral regions free of emission and sensitive upper limits on their emission can be derived. The remaining lines, at 221.55 and 222.96\,GHz, are expected to have peak intensities much below the rms noise and coincide with emission lines of TiO and SO$_2$. The four detected lines are included in Table\,\ref{Tab-TiO2}, and the unblended ones are shown in Fig.\,\ref{Fig-indiv-TiO2}. The integrated flux of the 221.1\,GHz line includes the contribution of the $^{34}$SO$_2$ emission and therefore is given as an upper limit in Table\,\ref{Tab-TiO2}.   

When the PdBI line profiles are compared to each other (Fig.\,\ref{Fig-indiv-TiO2}), the 220.8\,GHz line is centrally peaked at 30\,\kms, while the other two lines are dominated by blue-shifted components peaking between 5 and 10\,\kms. The profiles of the two latter lines are very similar. They have larger-than-average uncertainty in their rest frequencies and a relative shift of the order of one spectral channel ($\sim$2.7\,\kms) cannot be excluded. The emission in these two lines can be traced to {\it at least} --12\,\kms, which when combined with the positive velocity components observed in some of the SMA lines gives the full velocity range of TiO$_2$ emission between --12 and 62\,\kms\ -- see the comparison of the combined profiles\footnote{In the average PdBI profile, some emission outside the $-12\to 62$\,\kms\ range is seen, because the two averaged lines are not well separated in velocity.} in Fig.\,\ref{Fig-averprofile}. The PdBI data confirm that the broad emission of TiO$_2$ consists of several emission components that have different intensity ratios in individual transitions, suggesting that the components are characterized by a range of excitation temperatures. There is some correspondence (although not exact) in the positions of the main components seen in both combined profiles, e.g., both are dominated by the central double (U-shaped) peak. The PdBI spectra and measurements of TiO and TiO$_2$ provided in Tables\,\ref{Tab-TiO} and \ref{Tab-TiO2} and Figs.\,\ref{Fig-indiv-TiO} and \ref{Fig-indiv-TiO2} correspond to the full synthesized beam. 

\subsection{Isotopologues}\label{iso}
We did not observe lines of the rare isotopologues of TiO or TiO$_2$. Because the S/N of the main isotopic species is $<$26, non-detection of the rare isotopologues is consistent with the standard (solar and meteoritic) elemental composition in which the main isotope $^{48}$Ti is at least 9 times more abundant than all other isotopes of Ti. 


\subsection{SMA spatial constraints on the TiO and TiO$_2$ emission}\label{maps}

\begin{figure*}
\includegraphics[angle=270,width=0.325\textwidth]{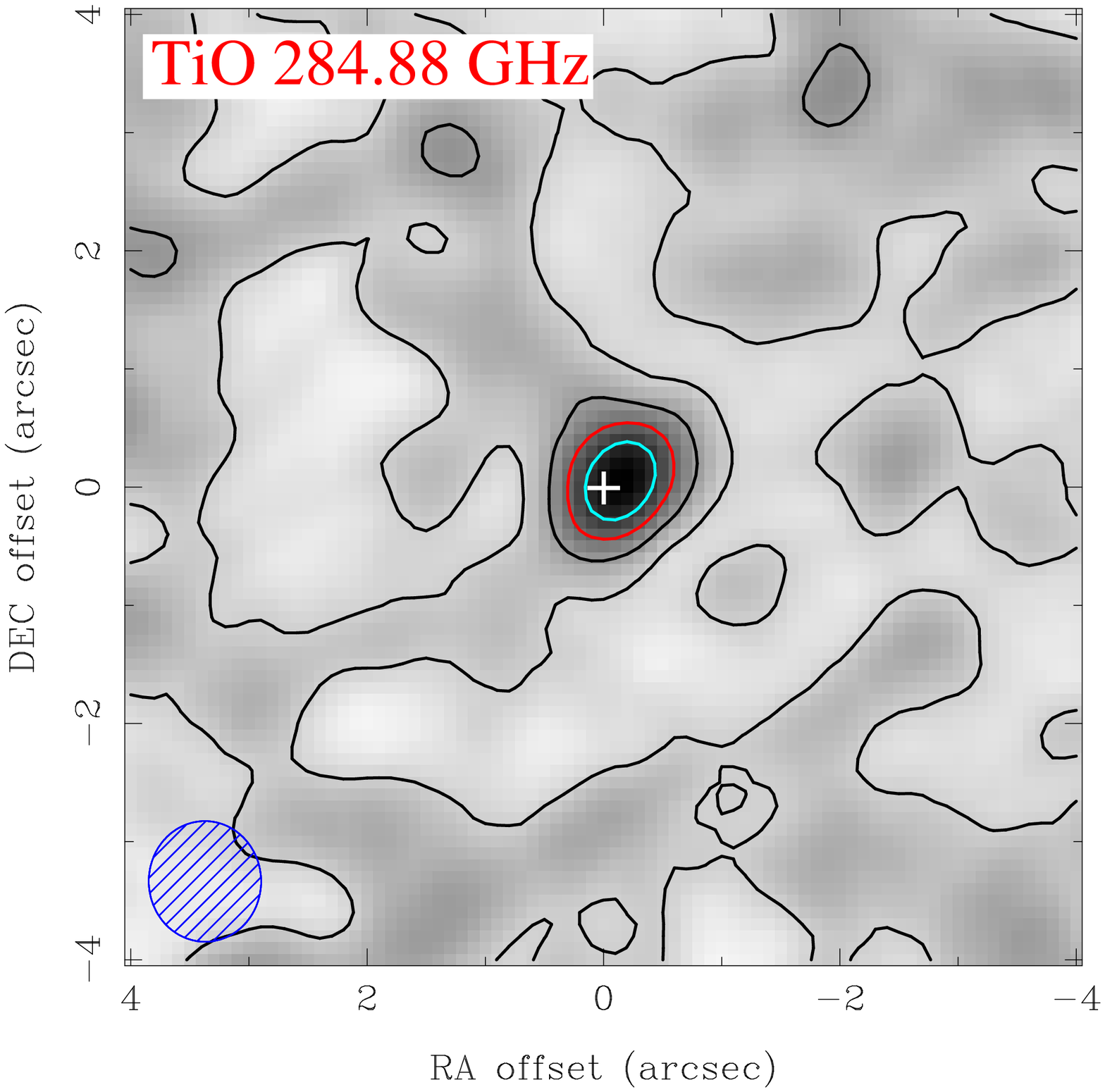}
\includegraphics[angle=270,width=0.325\textwidth]{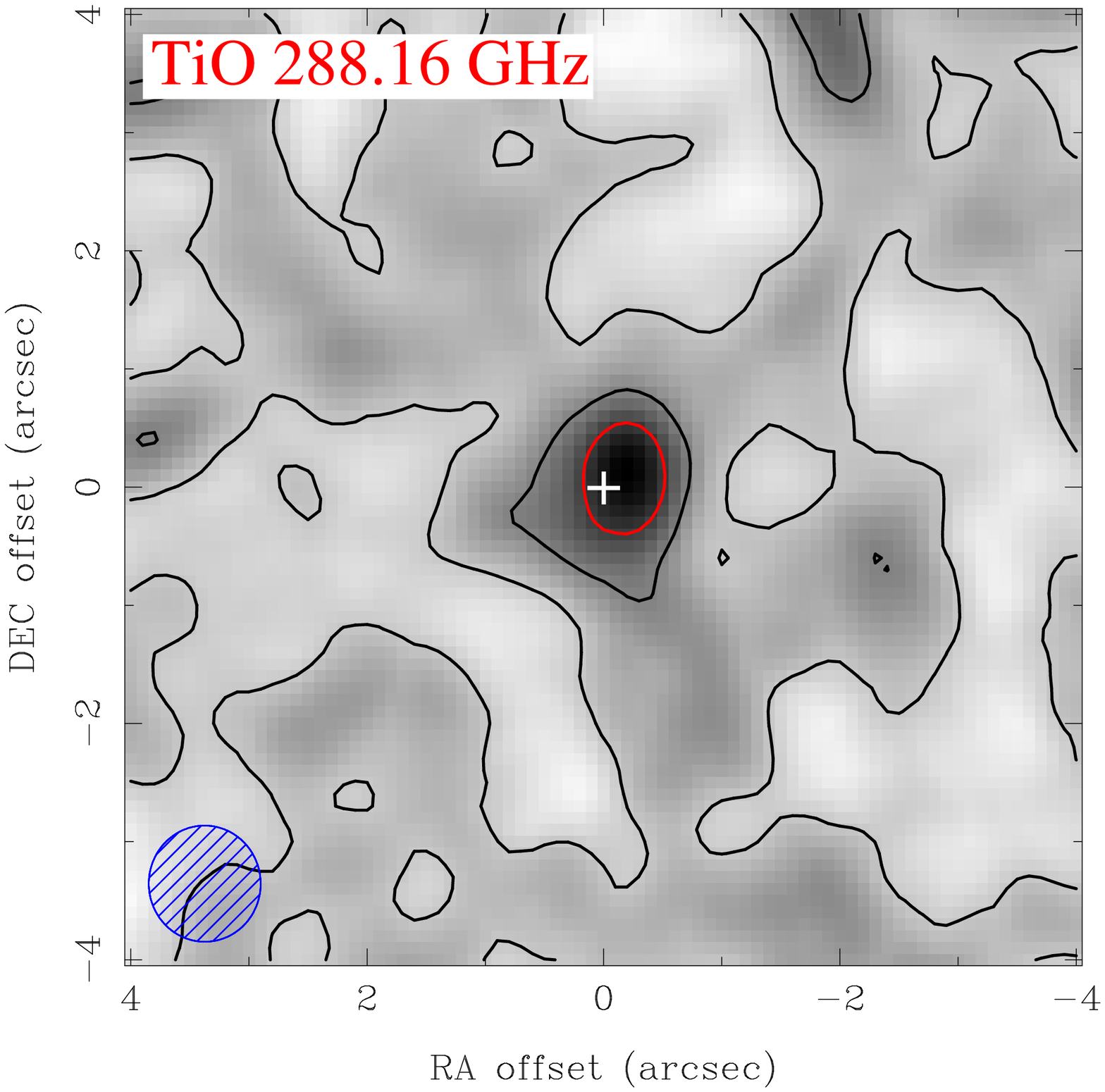}
\includegraphics[angle=270,width=0.325\textwidth]{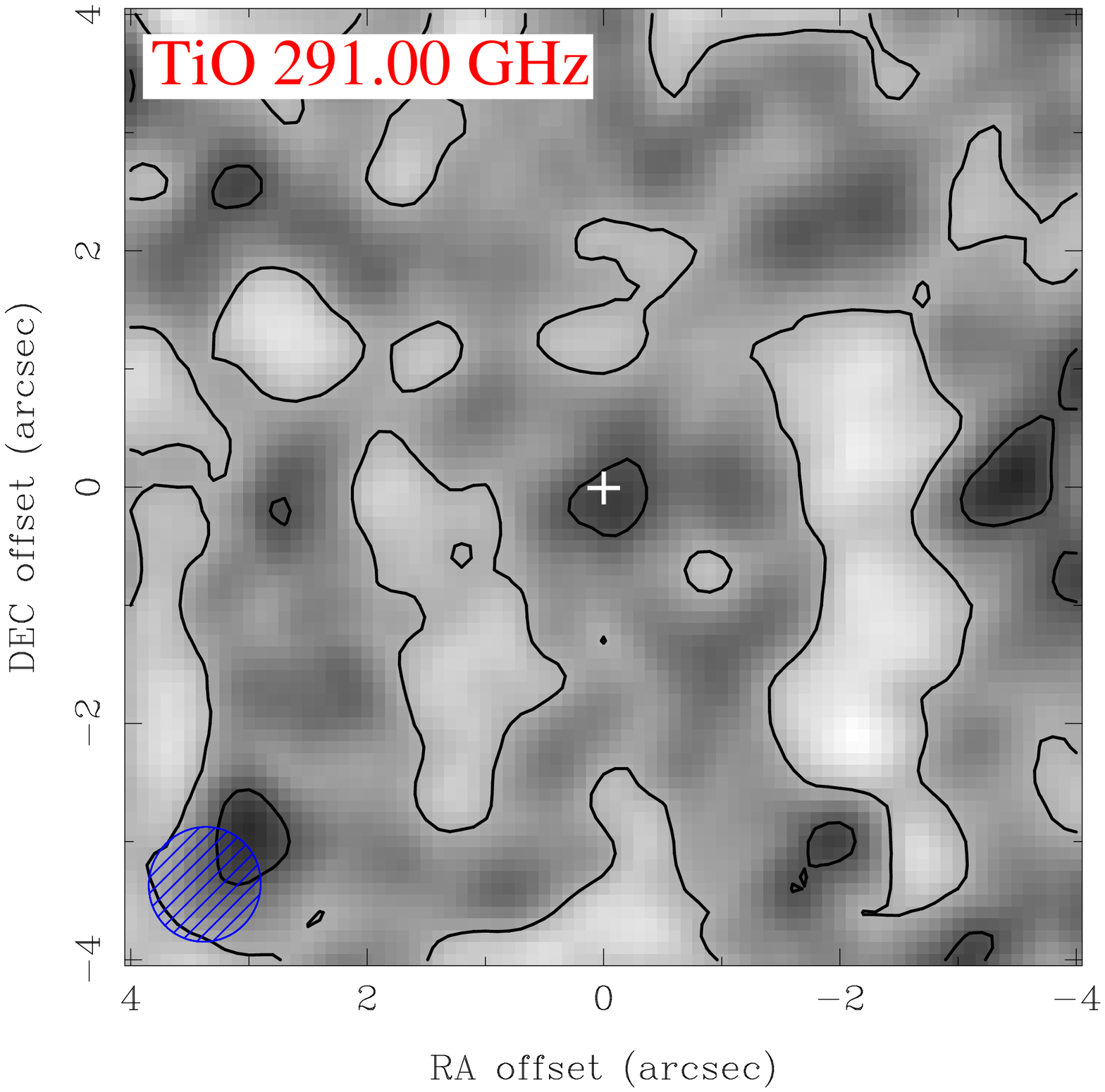}\\
\includegraphics[angle=270,width=0.325\textwidth]{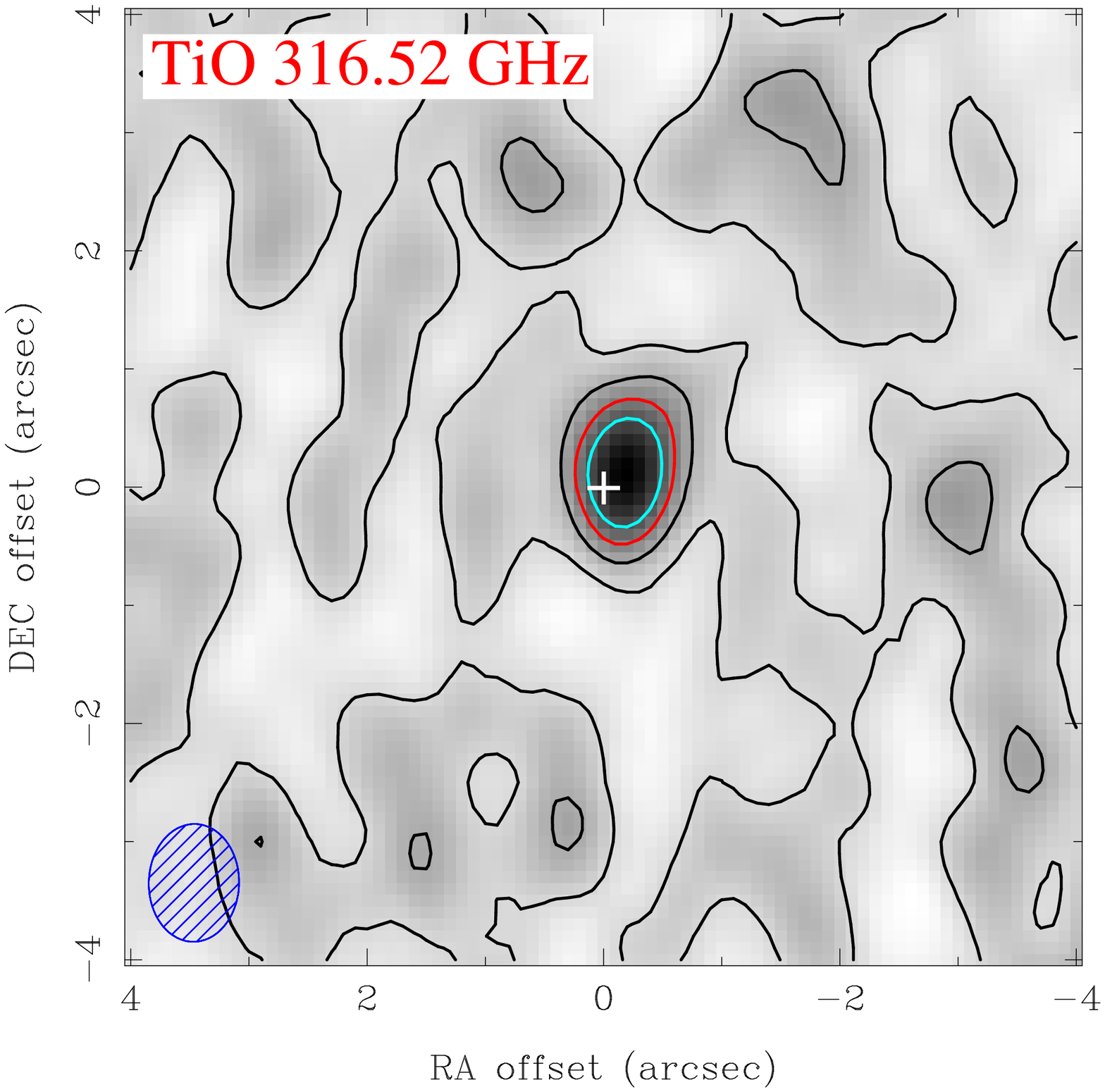}
\includegraphics[angle=270,width=0.325\textwidth]{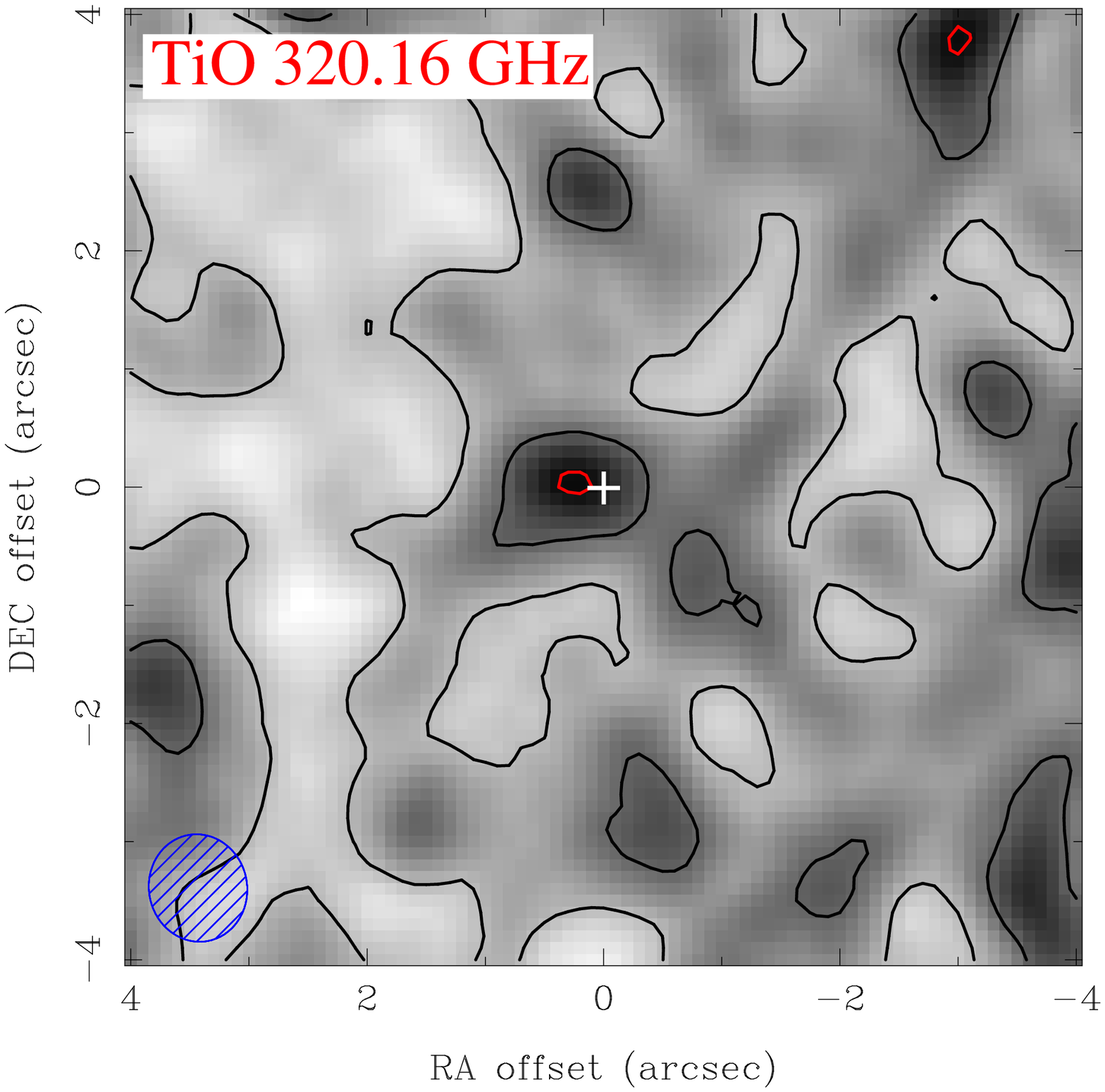}
\includegraphics[angle=270,width=0.325\textwidth]{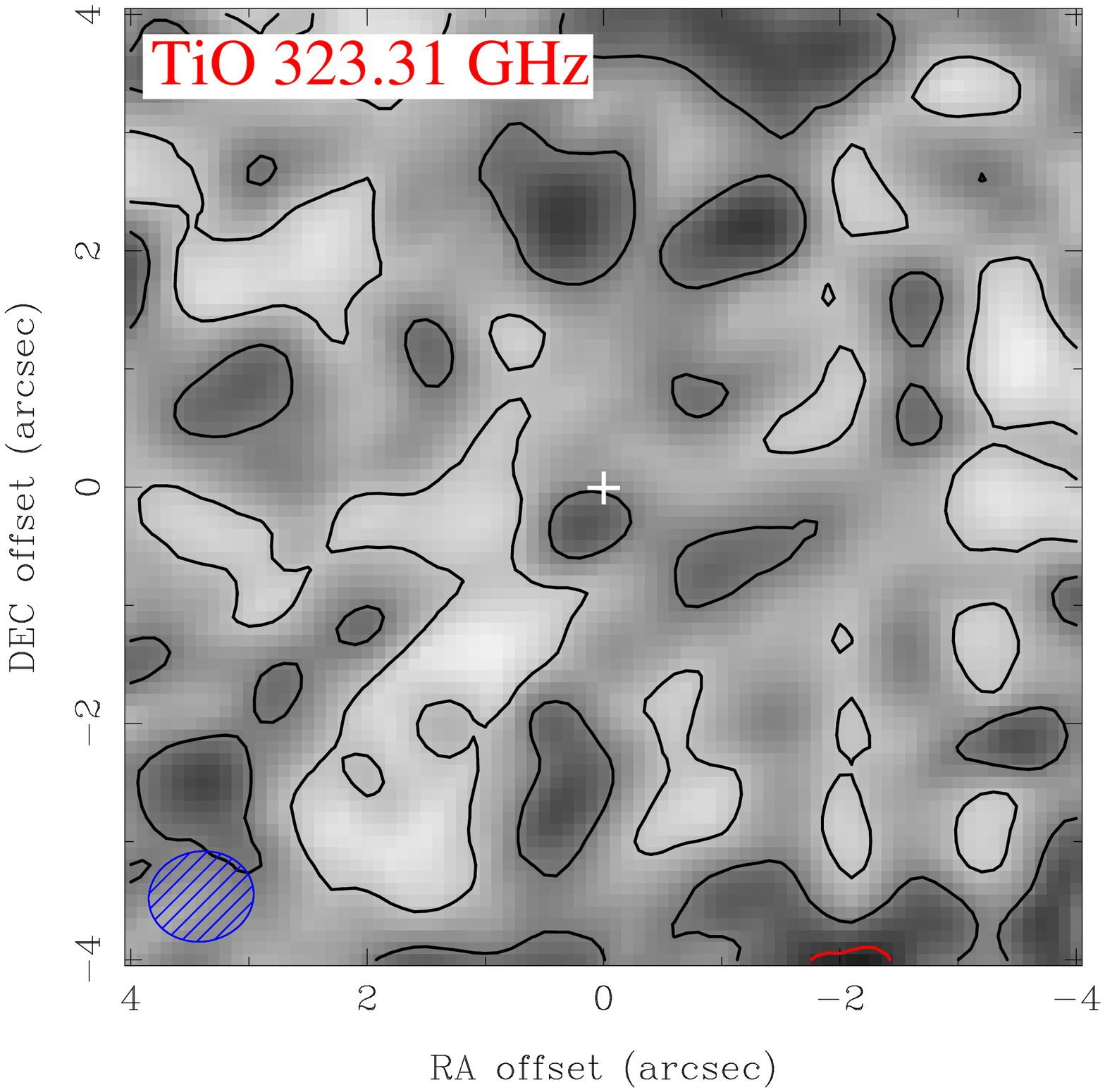}\\
\includegraphics[angle=270,width=0.325\textwidth]{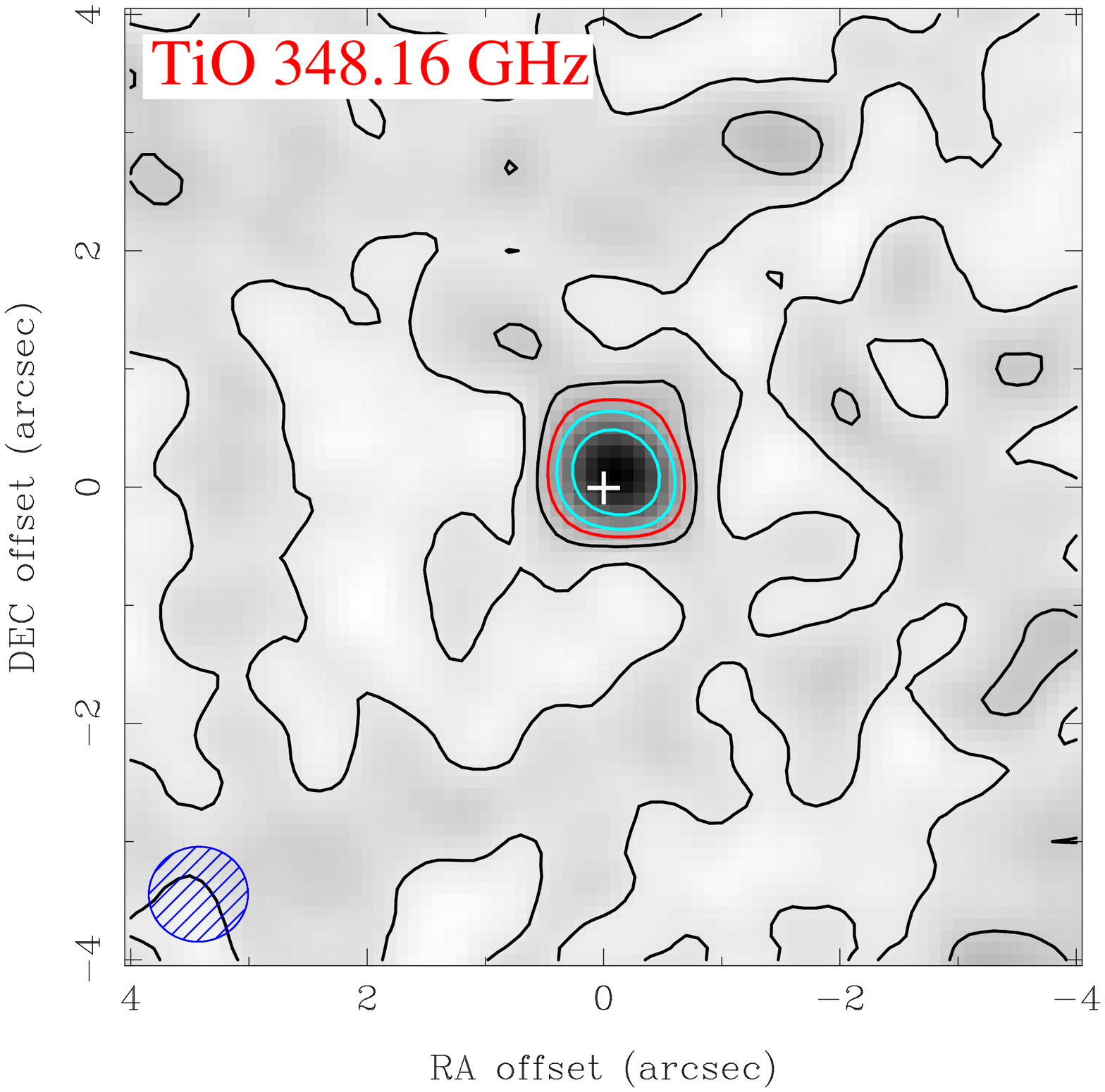}
\includegraphics[angle=270,width=0.325\textwidth]{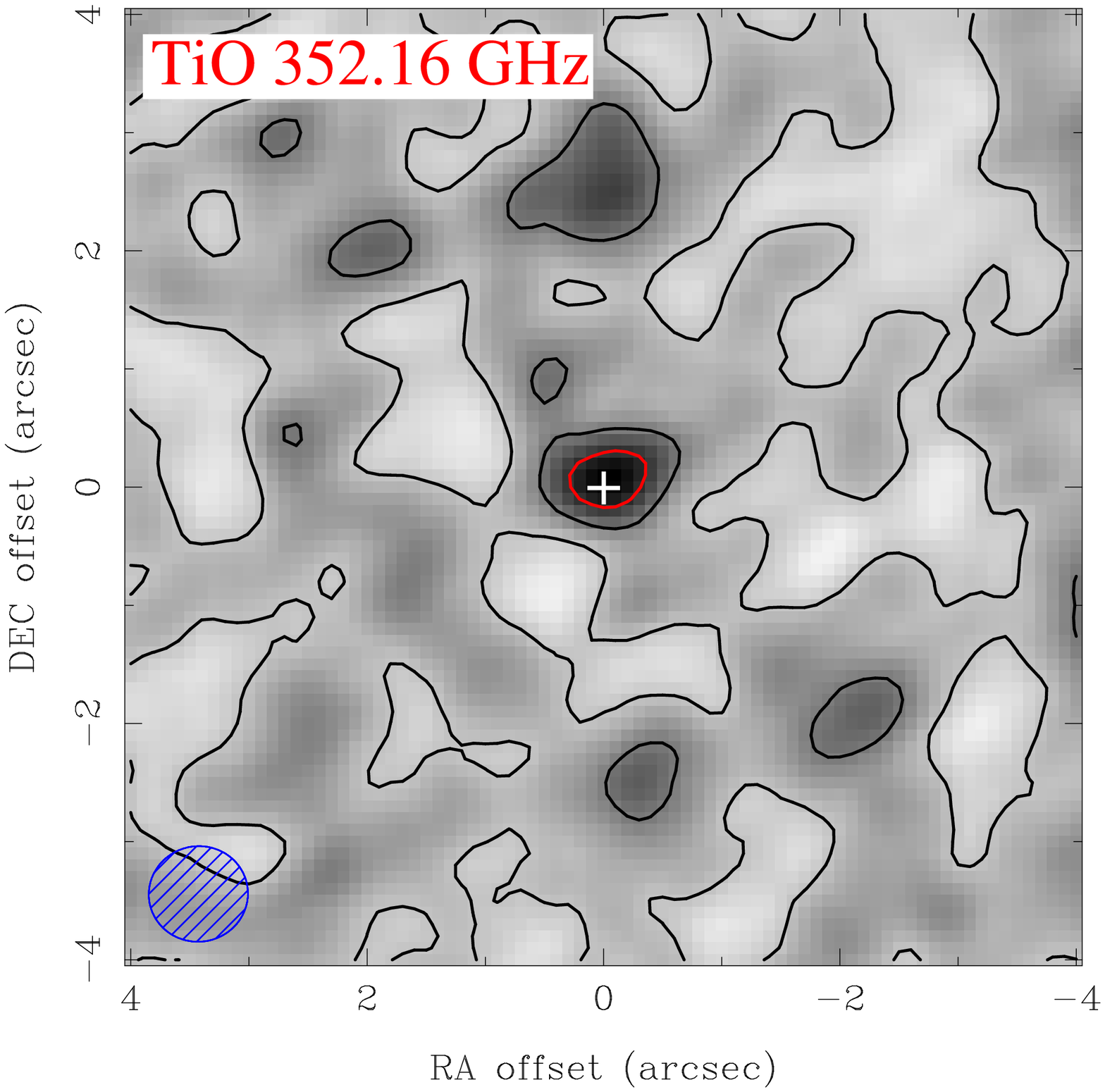}
\includegraphics[angle=270,width=0.325\textwidth]{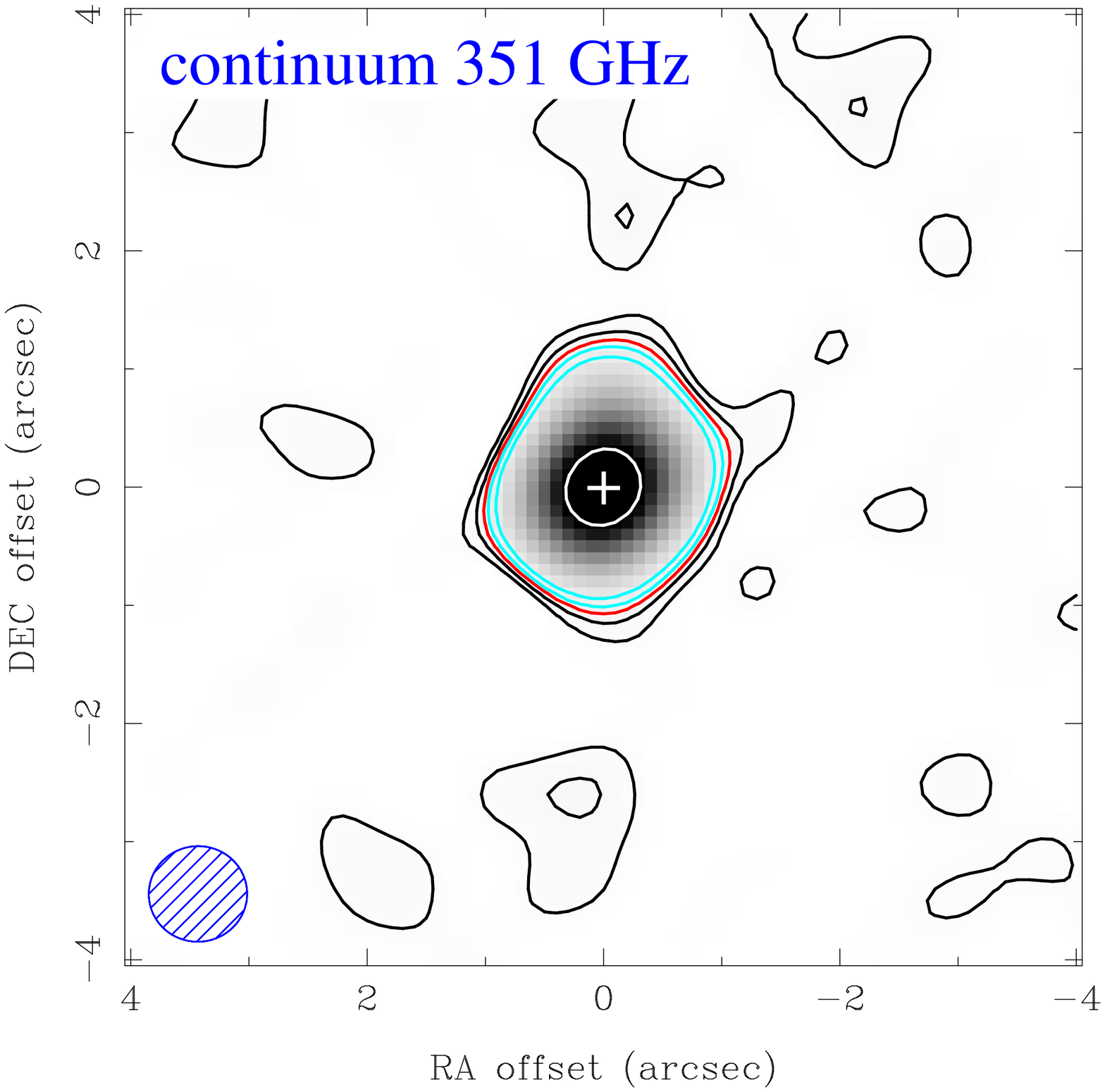}\\
\caption{All but the bottom right panel: Maps of emission integrated in the 6--36\,\kms\ range for all the TiO lines covered in the SMA survey. The columns are ordered (left to right) by $^3\Delta_1$, $^3\Delta_2$, $^3\Delta_3$; and the rows (top to bottom) by $J=9 \to 8$, $10 \to 9$, and $11 \to 10$. The bottom right panel: map of the continuum around 351\,GHz. The beam at half power is shown in the bottom left corner of each map. The contours are drawn at 1.5$\sigma$ and 3$\sigma$ (black), 4.5$\sigma$ (red), 6$\sigma$ and 9$\sigma$ (cyan), 120$\sigma$ (white), where $\sigma$ represents the noise levels measured in the integrated-intensity maps. The white cross (+) marks the peak of the continuum emission.}
\label{Fig-maps-TiO}
\end{figure*}

\begin{figure*}
\includegraphics[angle=270,width=0.325\textwidth]{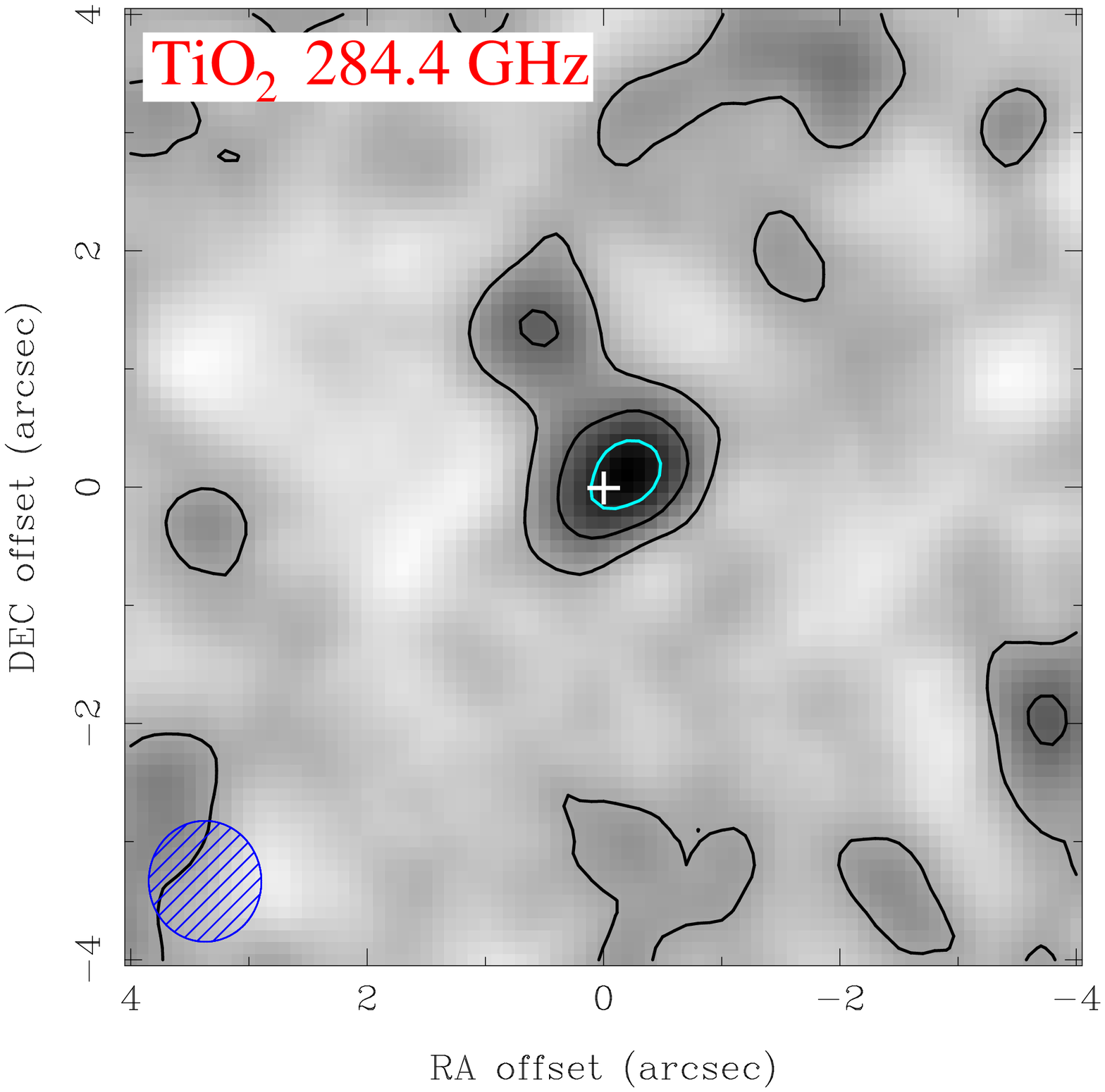}
\includegraphics[angle=270,width=0.325\textwidth]{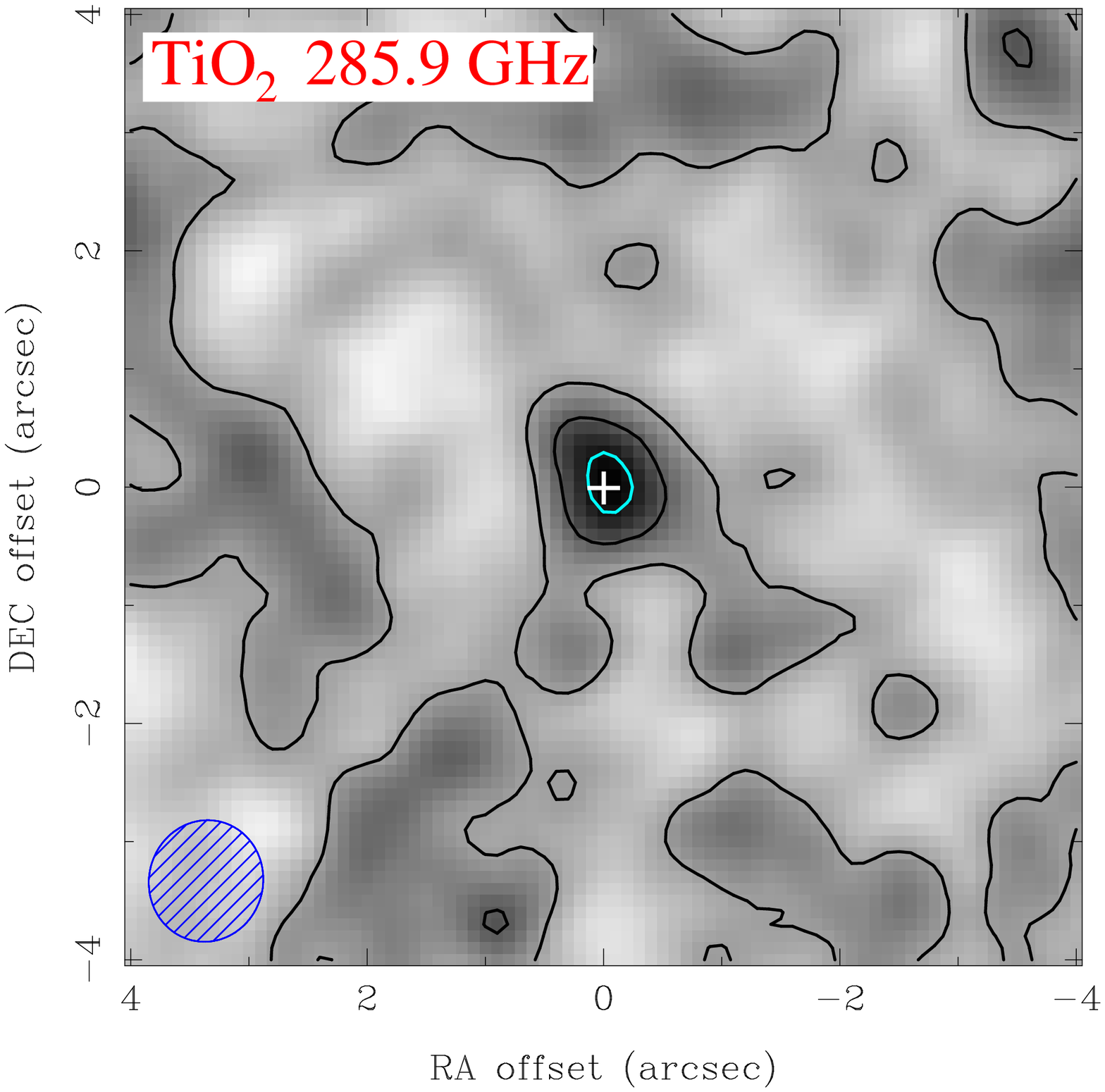}
\includegraphics[angle=270,width=0.325\textwidth]{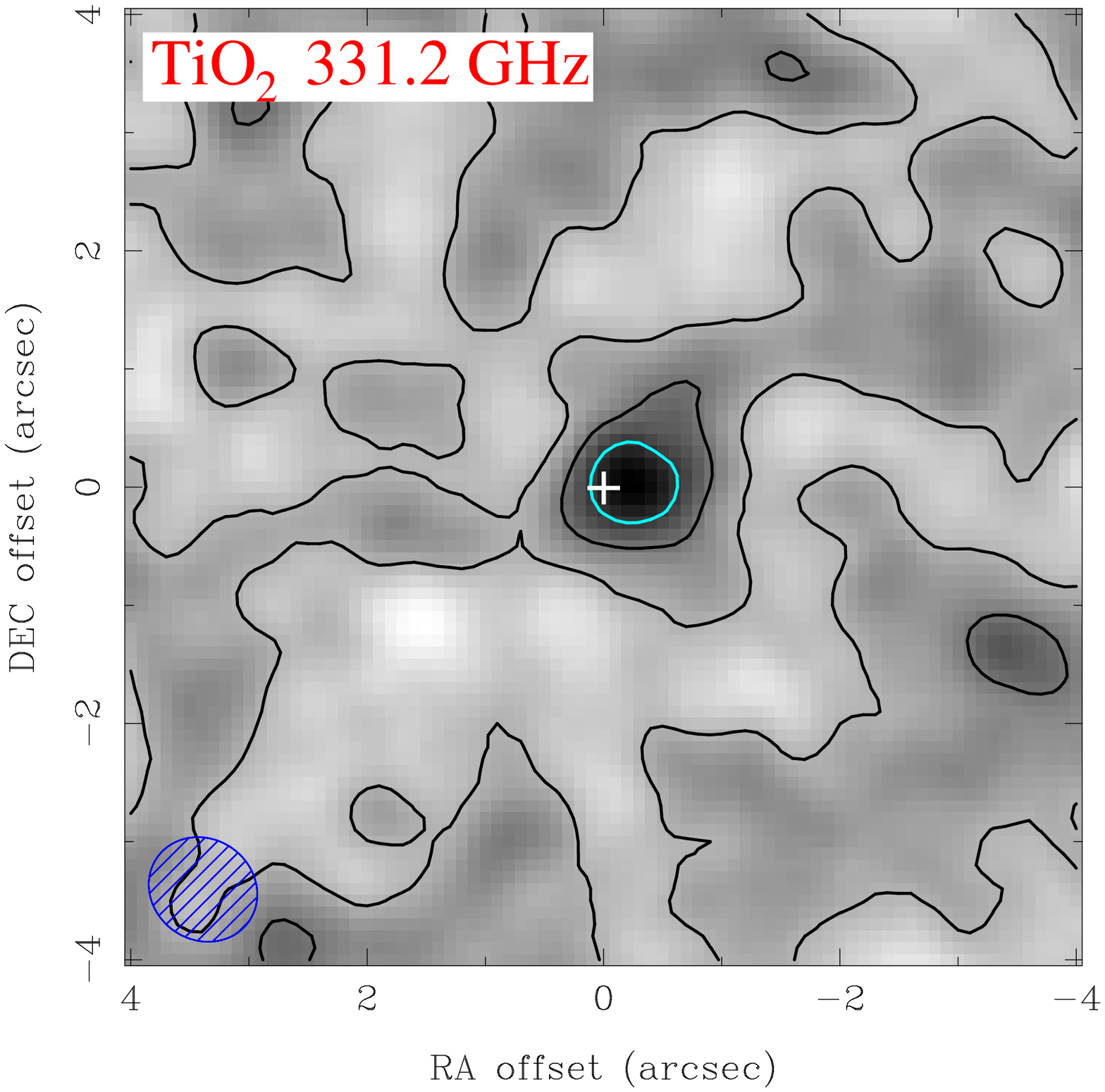}\\
\includegraphics[angle=270,width=0.325\textwidth]{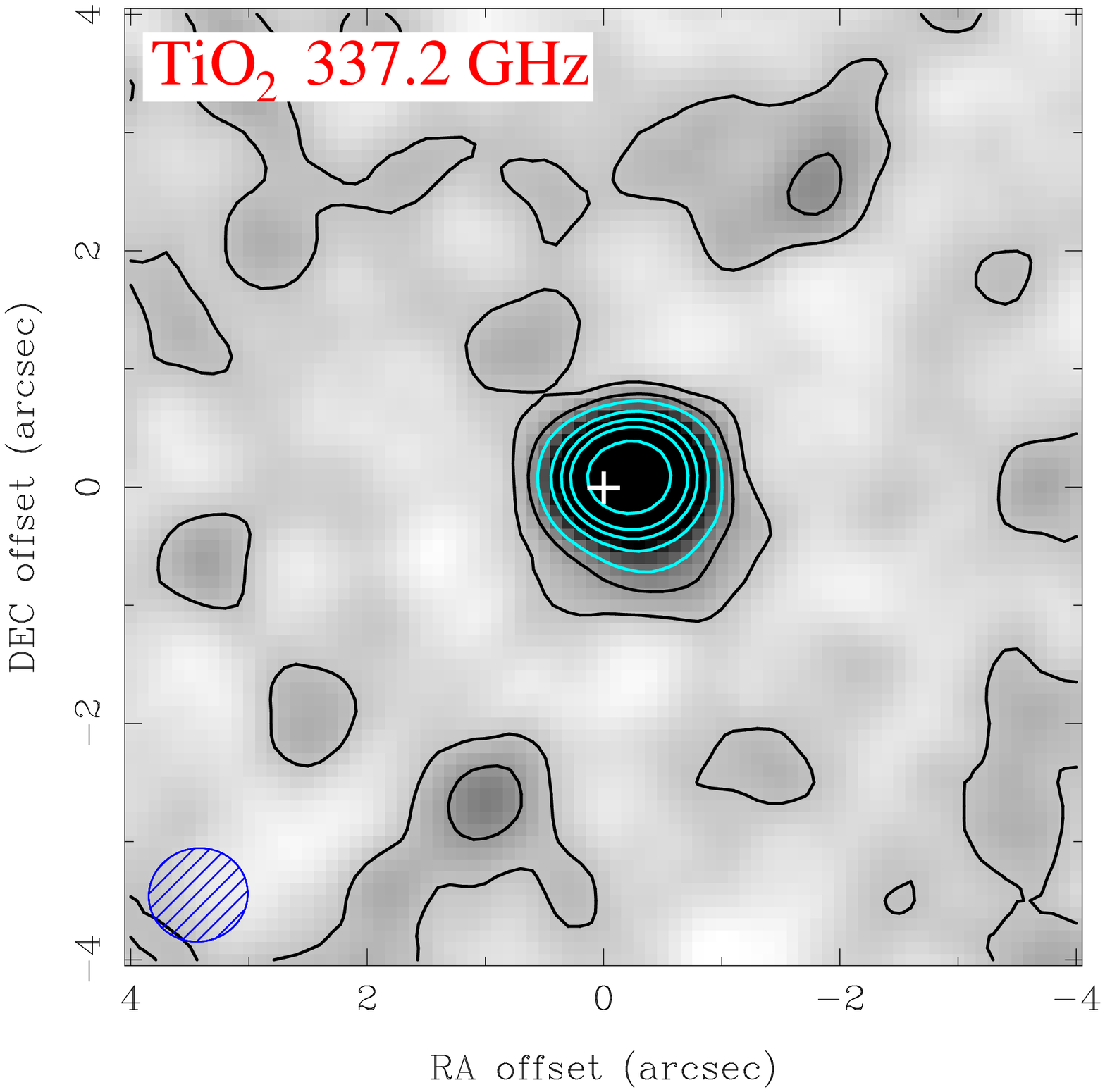}
\includegraphics[angle=270,width=0.325\textwidth]{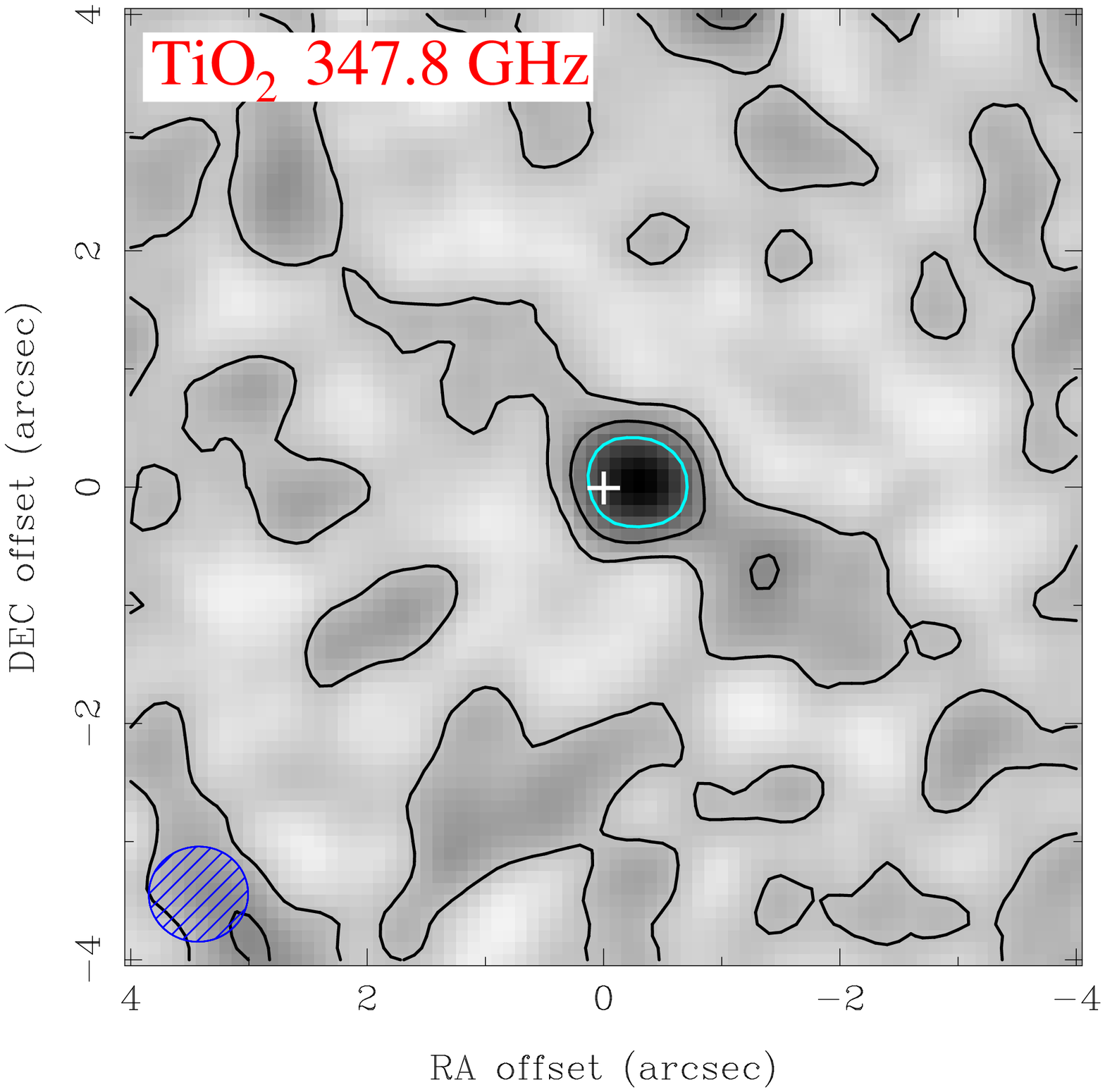}
\includegraphics[angle=270,width=0.325\textwidth]{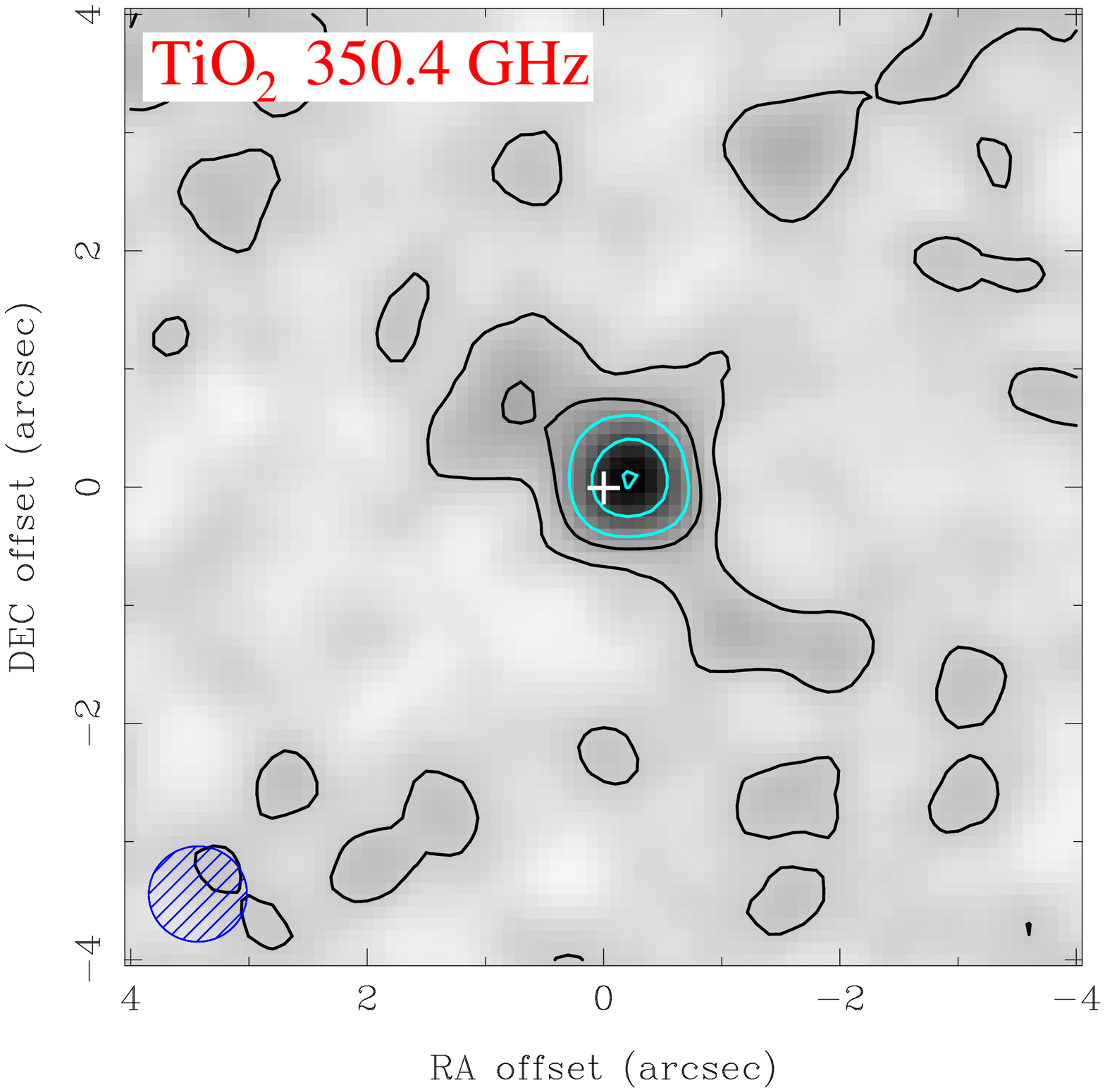}\\
\includegraphics[angle=270,width=0.325\textwidth]{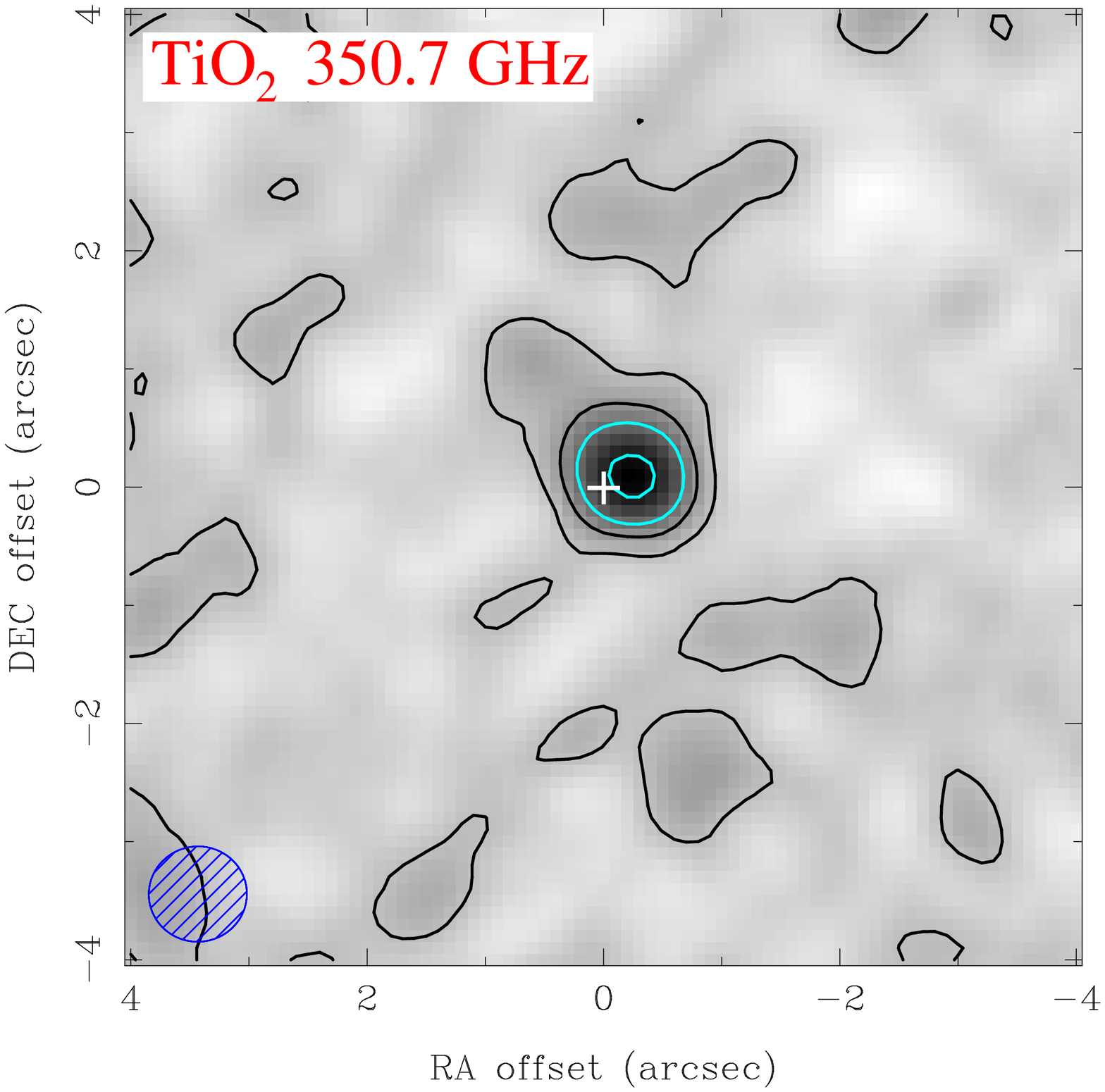}
\includegraphics[angle=270,width=0.325\textwidth]{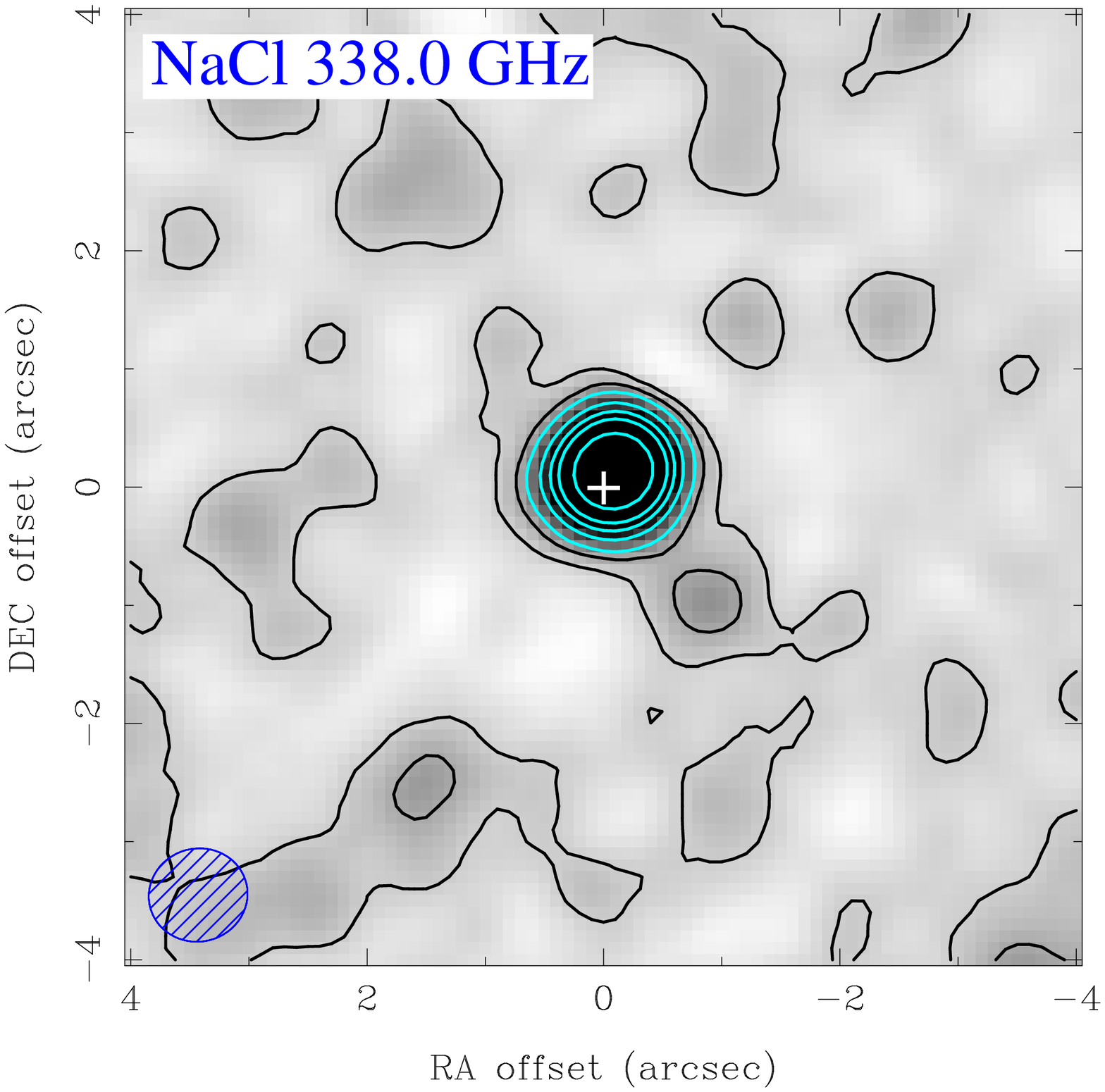}
\includegraphics[angle=270,width=0.325\textwidth]{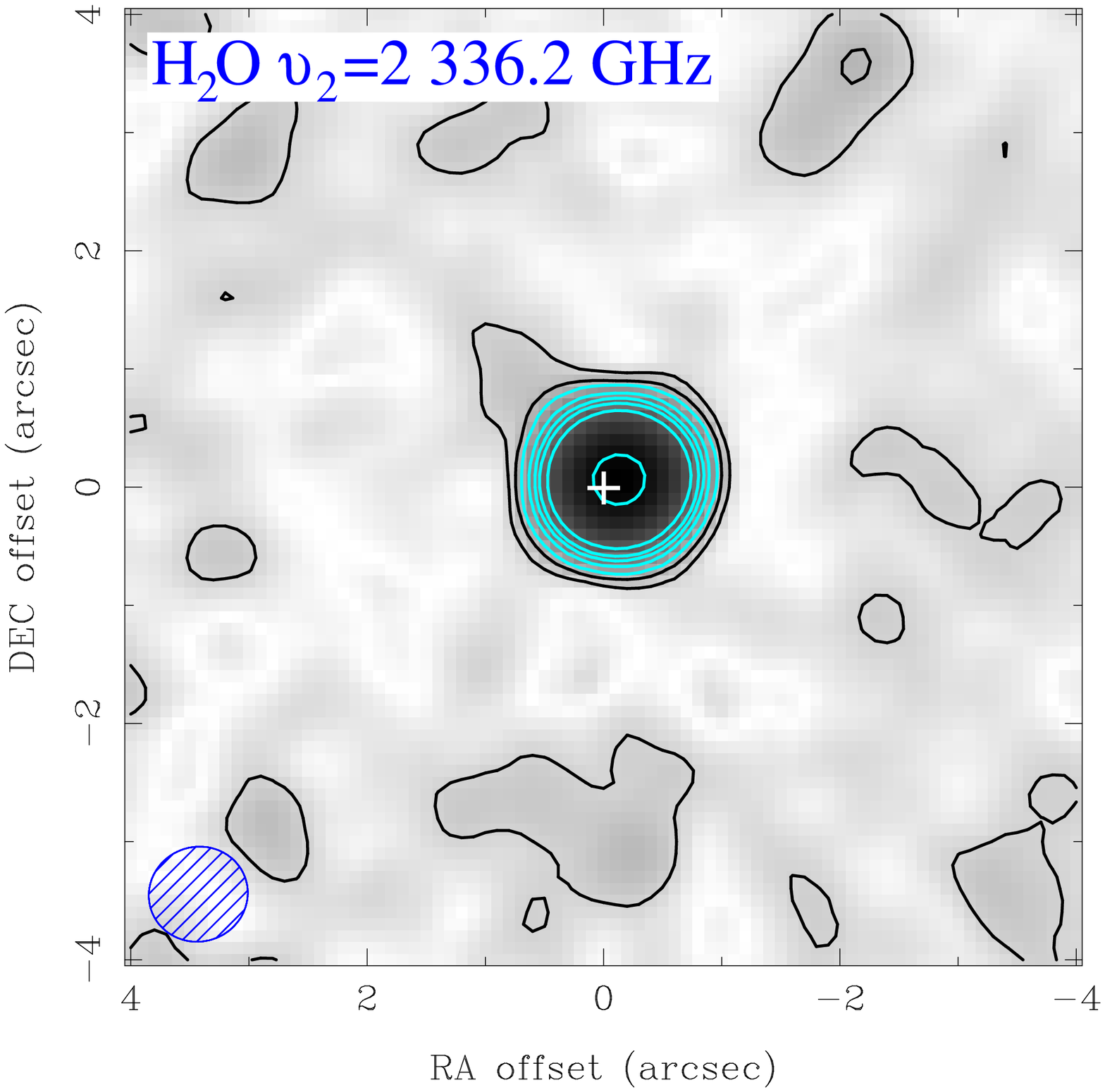}
\caption{The same as Fig.\,\ref{Fig-maps-TiO} but for the strongest emission features of TiO$_2$ and intensity integrated in the range 0--62\,\kms\ (adequately wider for blends). For comparison, emission of NaCl $J$=26$\to$25 and H$_2$O $\varv_2$=2 $J_{K_a,K_c}$=5$_{2,3}$\,$\to$6$_{1,6}$ is shown. The contours are at 5, 7.5$\sigma$ (black), and 10, 15, 20, 25, 35, 120$\sigma$ (cyan). See Table\,\ref{Tab-TiO2} for quantum numbers corresponding to the TiO$_2$ lines shown.}
\label{Fig-maps-TiO2}
\end{figure*}

Maps of integrated emission of TiO and TiO$_2$ are shown in Figs.\,\ref{Fig-maps-TiO} and \ref{Fig-maps-TiO2}. They are compared to maps of continuum emission at 351$\pm$2\,GHz, and that of NaCl $J$=26$\to$25 and H$_2$O $\varv_2=2$~$J_{K_a,K_c}$=$5_{2,3}$\, $\to$\,$6_{1,6}$, which are both expected to be compact sources \citep{milam,menten}. The center of the continuum emission is marked by a white cross in the figures. 

We measured the position and sizes of the TiO and TiO$_2$ emission by performing fits of a two-dimensional (elliptical) Gaussian to the emitting regions. If the synthesized beam is approximated by a Gaussian, then a deconvolved sizes of the emission can be calculated from the fits. For this purpose, we used an automated procedure {\it imfit} in Miriad. We found that for some TiO transitions the main emission component (between 10 and 30\,\kms) is a point source, i.e., its intrinsic size is much smaller than the beam. For a broader integration interval (6--32\,\kms), the average deconvolved size is 0\farcs5 FWHM. This value has however a large uncertainty owing to the very low S/N of the lines. For the TiO$_2$ emission analyzed within 0--62\,\kms, the average (weighted by the uncertainty of the Gaussian fit) deconvolved size is 0\farcs6$\pm$0\farcs1\ FWHM, in agreement with FWHM=0\farcs5$\pm$0\farcs1\ derived separately for the 337.2\,GHz feature alone, which has the highest S/N among all the lines analyzed.

Although the position of the emission for each individual TiO and TiO$_2$ feature usually agrees to within the (large) errors with the central position of the continuum source, in the case of the TiO line at 316.5\,GHz, the measured center is displaced by $-0$\farcs2\ in RA or 5 times the formal error of the fit; similarly, the strong TiO$_2$ emission at 337.\,GHz is displaced by $-0$\farcs2\ in RA or 4 times the error. The error-weighted mean values of the TiO and TiO$_2$ positions for all the analyzed features have a similar offset with respect to the continuum source: ($\Delta \alpha$, $\Delta \delta$)=(--0\farcs15$\pm$0\farcs06, 0\farcs12$\pm$0\farcs08) for TiO and (--0\farcs20$\pm$0\farcs05, 0\farcs08$\pm$0\farcs03) for TiO$_2$, where the 3$\sigma$ errors are indicated. The shift appears to be real and is distinguishable in Figs.\,\ref{Fig-maps-TiO} and \ref{Fig-maps-TiO2}. A similar offset is measured for other relatively compact molecular emission, e.g., that of vibrationally excited H$_2$O (see also NaCl; Fig.\,\ref{Fig-maps-TiO2}). The nature of the offset is discussed in Sect.\,\ref{dis1}.

The SMA measurements in Tables\,\ref{Tab-TiO} and \ref{Tab-TiO2}, and spectra in Figs.\,\ref{Fig-indiv-TiO}--\ref{Fig-indiv-TiO2} are for data extracted with a square aperture of 1\arcsec$\times$1\arcsec\ centered on the continuum peak. With the above constraints on the emission size and position, this aperture encloses more than 95\% of the emission thereby minimizing the contribution of noise in the spectra. We repeated the entire data analysis procedure including the possible shift in the extraction and obtained essentially the same results as those here.       
The PdBI measurements and spectra, correspond to the large synthesized beam of 4\farcs9$\times$2\farcs2, and therefore do not provide additional constraints on the size of the emission.


\subsection{Summary of evidence in support of the identification of TiO and TiO$_2$}\label{confi}

The evidence that we have observed pure rotational lines of TiO and TiO$_2$ in VY\,CMa is overwhelming. 
Following the initial identification of both molecules in the SMA survey, two new lines of TiO and four of TiO$_2$ were observed at the predicted intensities in a blind observation with PdBI. The agreement of the astronomical frequencies (V$_{\rm{LSR}}$) with laboratory measured frequencies, which are accurate to a small fraction of the line width, is excellent. There are no missing lines for either molecule, nor are there any lines that are anomalously intense. As shown in Fig.\,\ref{Fig-indiv-TiO}, the intensities of the eight lines of TiO between 221.58 and 352.16\,GHz that were detected and the two upper limits (at 291.00 and 323.31\,GHz), agree to within the measurement uncertainties with those derived from a simple LTE analysis (see Sect.\,\ref{rot-diagrams}). For TiO$_2$, approximately 20 unblended lines in the SMA survey with intensities within a factor of a few of the most intense lines at 350.4 and 350.7\,GHz, are present as predicted (Table\,\ref{Tab-TiO2}). Maps of the integrated emission of TiO in Fig.\,\ref{Fig-maps-TiO} confirm that all six lines observed with SMA are from a compact source (approximately) centered on the star, with an angular size that is much smaller than our 0\farcs9 synthesized beam, as expected for a species predicted to be localized in the inner outflow. 
As the sample maps in Fig.\,\ref{Fig-maps-TiO2} show, the emission from TiO$_2$ is similarly compact as that of TiO, and those of NaCl and vibrationally excited H$_2$O which were observed with the SMA under the same conditions and configuration as that of TiO and TiO$_2$.

Circumstantial evidence consistent with the identification of these two titanium oxides is provided by the line profiles, and an estimate of the likelihood of a misassignment owing to interfering background lines. Despite the fairly low S/N, the profile shapes of all the lines assigned to TiO and TiO$_2$ are consistent (Figs.\,\ref{Fig-indiv-TiO} and \ref{Fig-indiv-TiO2}). Those of TiO are narrow and centrally peaked, while lines of TiO$_2$  are broad and typically asymmetric with multiple velocity components.
Our confidence in the identification is further enhanced by the very low density of lines in our 77\,GHz wide survey of approximately 1~line every 380\,MHz. The average separation between lines in VY\,CMa is equivalent to about 25 linewidths of TiO, or 5 linewidths of TiO$_2$ --- i.e., there is very little of the overlapping of features that frequently occurs in sources which have a higher density of lines (e.g., IRC+10{\degr}216 and hot cores). Therefore, the probability that the emission features assigned to all 8 lines of TiO, or all 27 lines of TiO$_2$ are due to chance coincidences with lines of other species is exceedingly low in VY\,CMa: $< 10^{-10}$ for TiO and $< 10^{-18}$ for TiO$_2$.\footnote{For TiO$_2$, the probability is even lower as more than 27 lines are seen. Above, we have presented only lines that are not contaminated by any other emission, while there are many examples of TiO$_2$ lines which interfere with emission lines of other species; the presence of TiO$_2$ in many such cases is evident when the blended profile is compared to a sequence of other uncontaminated profiles of the species.}

The velocity ranges in which the emission of TiO and TiO$_2$ is observed are well within those of other firmly identified molecules in VY\,CMa. For example, in Fig.\,\ref{Fig-averprofile} we compare the newly discovered emission to a profile of NaCl combined of many lines detected in the survey, and to the strongest emission line of SO$_2$ (at 283.46\,GHz), all extracted for the same aperture 1\arcsec$\times$1\arcsec. Lines of NaCl are among the narrowest features observed in VY\,CMa \citep{milam}, while SO$_2$ shows broad, multicomponent emission typical of many sulphur-bearing species in this object. The weak emission of the two titanium oxides is located almost exactly within the velocity range of the SO$_2$ profile, while the main component of the TiO emission covers the same range as the NaCl profile. This correspondence is additional strong evidence in support of our identification. 


\section{Analysis of rotational-temperature diagrams }\label{rot-diagrams}
We made an attempt to constrain the physical conditions of the emitting gas by analysing the Boltzmann (rotational temperature) diagrams \citep[e.g.,][]{snyder} for both molecules under the assumption of LTE and optically thin emission. The rotational temperature ($T_{\rm rot}$) and total column density ($N$) were obtained from results of least-squares fits to the measurements weighed by their uncertainties. The fluxes of the PdBI data were scaled to correspond to the aperture of 1\arcsec$\times$1\arcsec. 

\begin{figure}\centering
\includegraphics[angle=270,width=\columnwidth]{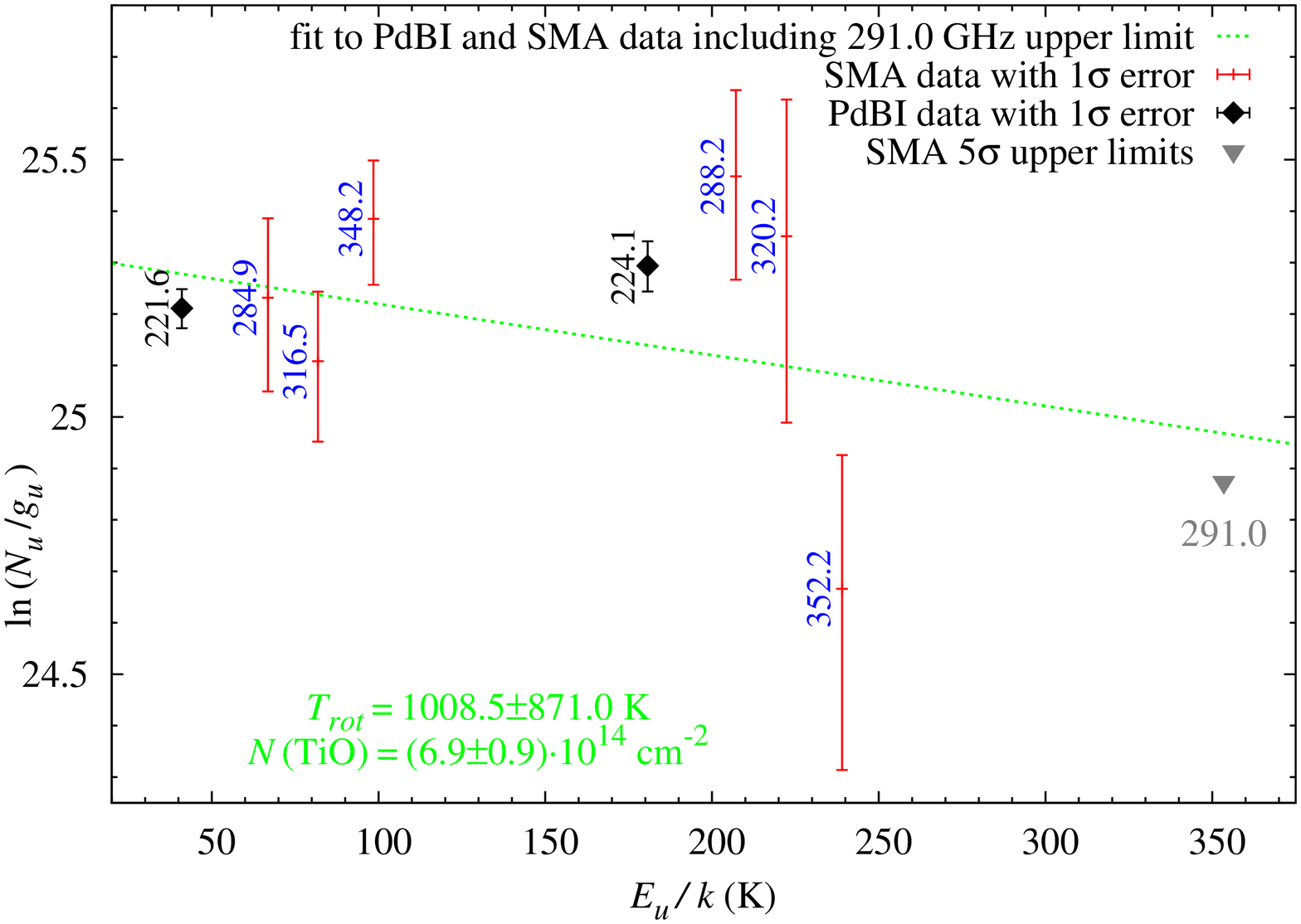}
\caption{Rotational temperature diagram of TiO. The upper limit on the line at 323.31\,GHz is above the range on the vertical scale.}
\label{Fig-rotdiagr-TiO}
\end{figure}
 
{\bf TiO:} Shown in Fig.\,\ref{Fig-rotdiagr-TiO} is a rotational temperature diagram of TiO derived  from measurements in Table\,\ref{Tab-TiO}. Owing to the large uncertainties in the fluxes of the eight lines because of noise and systematic effects, we were unable to constrain the rotational temperature when the line at 291.0\,GHz was omitted in the analysis -- i.e., the line representing the best fit parameters has a negative slope which is nonphysical under LTE conditions and inconsistent with the upper limit on the 291.0\,GHz line. As mentioned in Sect.\,\ref{ana1}, the line at 291.0\,GHz is close to the detection limit. When the upper limit on this  emission was taken into account in the fitting procedure, we obtained $T_{\rm rot}$=1010$\pm$870\,K and $N({\rm TiO})$=(6.9$\pm$0.9)$\cdot10^{14}$\,cm$^{-2}$. This solution is shown in Fig.\,\ref{Fig-rotdiagr-TiO} and also in Fig.\,\ref{Fig-indiv-TiO}, where Gaussian profiles are overlaid on observed lines to illustrate the relative intensity scale at the derived temperature.  The quoted errors were derived from the formal uncertainties of the fit. While the rotational temperature remains poorly constrained, the uncertainty in column density is small. 

The failure to constrain $T_{\rm rot}$ from the diagram may imply that the assumption of LTE is not valid for the TiO gas. Indeed, VY\,CMa displays optical TiO emission whose presence implies non-LTE effects, most likely related to the strong stellar radiation \citep[cf.][]{kami_alo}. Large scatter of points on a rotational temperature diagram and inability to constrain $T_{\rm rot}$ are problems we have encountered for several other molecules observed in the SMA survey, including NaCl and SiS for which many lines ($>$15) are observed, all at higher S/N. Although, optical and infrared pumping may be important in the rotational excitation of TiO, the uncertainties in the current measurements are too large to allow any conclusive statements. 

\begin{figure}\centering
\includegraphics[angle=270,width=\columnwidth]{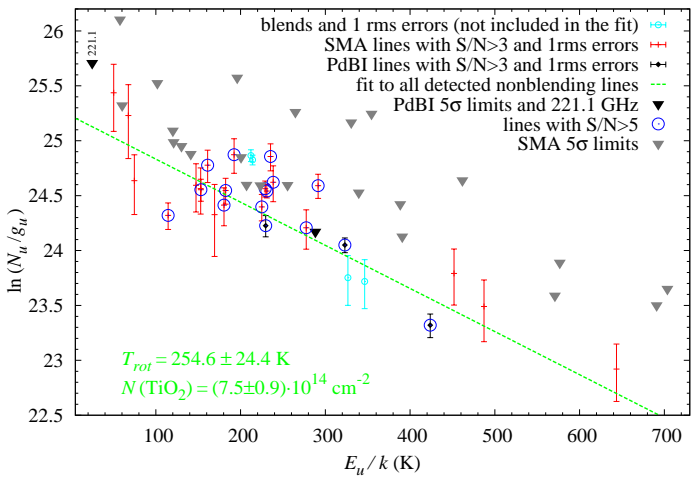}
\caption{Rotational temperature diagram of TiO$_2$.}
\label{Fig-rotdiagr-TiO2}
\end{figure}

{\bf TiO$_2$:} A rotational temperature diagram of TiO$_2$ derived from the measurements in Table\,\ref{Tab-TiO2} is shown in Fig.\,\ref{Fig-rotdiagr-TiO2}. For those lines that are blended and arise from levels of similar energy, the flux of the individual line was estimated from the intensity of the observed feature divided by the Einstein coefficient ($A_{ul}$) of each line (see the lower part of Table\,\ref{Tab-TiO2} labelled ``SMA deblended''). The deblended lines are plotted in the diagram but were not included in the fit. The best fit to all detected lines (SMA and PdBI) gives $T_{\rm rot}$=255$\pm$24\,K and $N$(TiO$_2$)=(7.5$\pm$0.9)$\cdot$10$^{14}$\,cm$^{-1}$. This solution is also consistent with numerous 5$\sigma$ upper limits, some of which can be seen in the diagram. 

Although the fit is satisfactory (the confidence level is above 99\%), it corresponds to the {\it average} gas parameters over the entire velocity range between 0 and 62\,\kms\ and omits the most blue-shifted component in the PdBI spectra. The different velocity components contributing to the TiO$_2$ emission may have different temperatures.  

In the above analysis, the data from SMA and PdBI were combined despite the possibility that the molecular emission is time variable. We note,  however, that temperatures calculated for PdBI and SMA data separately are consistent within uncertainties ($T_{\rm rot}^{\rm PdBI}$=200$\pm$60\,K and $T_{\rm rot}^{\rm SMA}$=305$\pm$35\,K, respectively), justifying our fit to the combined data.

\section{TiO at radio and optical wavelengths}\label{optical}

\begin{figure}\centering
\includegraphics[angle=270,width=\columnwidth]{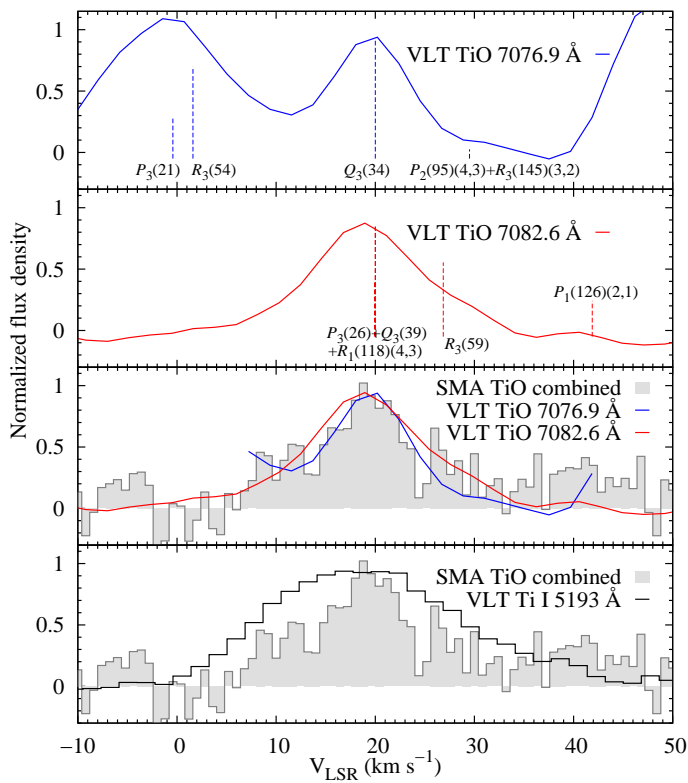}
\caption{Rotational profiles of TiO emission in the $\gamma$ system near 7080\,\AA\ (panels first and second from the top) are compared to the combined SMA profile of TiO (third panel from the top). Bottom panel: the \ion{Ti}{I} line at 5193\,\AA\ is overlaid on the combined SMA profile. The vertical lines mark positions of rotational components of the (0,0) band, unless otherwise indicated in the labels in the Figure. All the optical lines were shifted arbitrarily by --2\,\kms\ to facilitate direct comparison with the SMA radio profile.}
\label{Fig-optical-comp}
\end{figure}

To compare the rotational profiles of TiO observed at radio and optical wavelengths, we used an optical spectrum of VY\,CMa obtained in 2001 with the Very Large Telescope (VLT) and reported in \citet{kami_alo}. Refering to the TiO line list of G.\,J.\,Phillips\footnote{available on the web site of R.\,L.\,Kurucz http://kurucz.harvard.edu}, we searched for single-component and well separated rotational lines, and chose two examples in the $\gamma$ (0,0) band. A line at the rest wavelength $\lambda_{\rm lab}$=7076.93\AA\ is mainly formed by the $Q_3$($J$=34) transition and is affected by the presence of components from higher vibrational bands (at 9\,\kms\ with respect to $\lambda_{\rm lab}$); moreover, the wings of the line are blended with those of other nearby rotational lines. Another line, at $\lambda_{\rm lab}$=7082.64\AA, is formed mainly by $P_3$(26) and $Q_3$(39) (and $R_1$(118) of the (4,3) band) and is much better separated from the neighboring rotational components; its red wing, however, has some contribution from the weak $R_3$(59) line (at 7\,\kms\ with respect to the main components). The structure of the lines is illustrated in Fig.\,\ref{Fig-optical-comp}, where the profiles are also compared to the combined SMA profile.  

The profile of the optical and radio emission is very similar. The optical lines are however slightly narrower than the combined SMA profile if results of Gaussian fits performed within 3--32\,\kms\ are compared; the average FWHM of the two optical lines shown in Fig.\,\ref{Fig-optical-comp} is 12$\pm$3\,\kms. Some other rotational lines measured are as narrow as 9\,\kms\ FWHM, but due to unknown level of the photospheric fluxes these widths may be underestimated. For the same reason, the presence of the broad pedestal seen in the radio spectra cannot be investigated. The peaks of the optical lines are at $V_{\rm LSR}$=22$\pm$1\,\kms\ and there is no systematic shift of line centers with $J$. Taking into account the uncertainty in the wavelength calibration of 3\,rms=0.7\,\kms, this peak velocity is formally consistent with the location of the center of the combined SMA profile ($V_{\rm LSR}$=20$\pm$2\,\kms). Similarly, \citet{waller86} derived $V_{\rm LSR}$=24$\pm$2\,\kms\ from positions of the TiO band heads, which to within the errors is consistent with the peak position of the radio emission. However, when the profiles are overplotted, the optical lines appear to be systematically shifted by about 2\,\kms\ with respect to the radio profile (note that a compensating correction was applied in Fig.\,\ref{Fig-optical-comp}). Some red-shift of the optical emission with respect to the radio emission is expected owing to scattering of the optical radiation on grains within the dusty nebula \citep{vanBlerkom,humphreys05}. As noted in \citet{kami_alo}, this red-shift of optical lines must be less than about 3\,\kms\ and does not affect considerably the profiles for the line of sight towards the star; both effects are significant only in the outer parts of the nebula \citep{humphreys05}. 

Rotational temperatures derived from our measurements of the pure rotational spectrum and previous optical studies of TiO are consistent, although the uncertainties are high in both. \citet{phillips} derived a rotational temperature of 600\,K with an uncertainty of about 150\,K from the rotational structure of the $\gamma^{\prime}$ system at 6200\,\AA~(B$^3\Pi$--X$^3\Delta$), however there are two possible sources of systematic error in the optical measurements.    
First, the analysis of \citet{phillips} requires a revision with a new TiO line list containing improved oscillator strengths \citep[e.g.,][]{schwenke}. The new data can change the shape and intensity of the simulated spectrum of the emission (and the underlying photospheric absorption), thus yielding revised estimates of the rotational temperature \citep{allard}. 
Second, because the optical emission spectrum of VY\,CMa and the bands of TiO are time variable \citep[e.g.,][]{waller86}, the excitation temperature of TiO may be time variable too. For example, rotational temperatures derived from the orange (0,0) band of the A$^2\Pi$--X$^2\Sigma$ system of ScO differed by more than a factor of two on two different dates ($T_{\rm{rot}}$=380\,K and 820\,K in 1962 and 1966, respectively; \citealp{herbig74}). 

The above comparison provides further strong evidence in support our identification of TiO radio emission. The similarity of the optical profiles and the main (central) component of the (sub-)millimeter emission suggests that they arise in the same circumstellar gas. 
\section{Discussion}\label{dis}
\subsection{Localization and abundances of the TiO and TiO$_2$ gas}\label{dis1}

The displacement of the center of the molecular emission with respect to continuum peak (Sect.\,\ref{maps}) brings into question the actual position of the star in our self-calibrated maps. The continuum source is resolved in the SMA observations at a very high S/N ($\sim$200 for the peak flux in a single 2\,GHz-band) and is elliptical in shape with 0\farcs5$\times$0\farcs3 FWHM. Therefore the origin of the radio continuum is certainly not entirely photospheric. Indeed, models of the spectral energy distribution of VY\,CMa \citep[e.g.,][]{sed} suggest that at the observed frequencies the total flux is almost entirely reprocessed dust radiation, with the direct photospheric flux constituting only a few percent of the total emission. Our measured flux of 0.6\,Jy at 320\,GHz is consistent with those models. Although both dust and molecular gas are likely to have a very inhomogeneous distribution around VY\,CMa even at lowest spatial scales, we believe it is the center of molecular emission that can be identified as the position of the star itself. In support to this interpretation comes the observation that emission of transitions from highly excited states, like that of  H$_2$O $\varv_2$=2 $J_{K_a,K_c}$=5$_{2,3}$\,$\to$6$_{1,6}$ with $E_u$=2955\,K, peaks at this location; since it requires a nearby energetic excitation source, e.g., energetic radiation or high-density gas of high temperature, it is very likely located close to the star. Additionally, to produce centrally peaked profiles (like those of TiO and NaCl) and emission center displaced with respect to the kinematical center, the gas would have to move tangentially with almost no radial component, which is very unlikely in this source. Moreover, \citet{muller} noticed a very similar spatial shift between the centroid of the continuum emission at 1.3\,mm and the positions of the masers of SiO, H$_2$O, and OH in VY\,CMa. The positional shift between the continuum and molecular emission will be discussed in more detail in a separate paper. In the following discussion we assume the observed molecular emission is spatially centered on the star.  

That the TiO$_2$ emission reaches the outflow terminal velocity \citep[$V_{\infty}$=40--45\,\kms,][]{humphreys05} implies that at least part of the outflow is already accelerated to $V_{\infty}$. In case of TiO, the broad pedestal can be traced to high expansion velocities but due to a very low S/N it is uncertain whether it indeed reaches $V_{\infty}$ (Fig.\,\ref{Fig-averprofile}). The main Gaussian-like component of the TiO emission, however, is very likely formed in the inner outflow because it is centrally peaked and much narrower than 2$V_{\infty}$. Similar narrow molecular emission lines in VY\,CMa have been attributed to the accelerating wind \citep[e.g.,][]{milam}. Following \citet{tanen_alo} and \citet{kami_alo}, to better localize the TiO emission region we use the semi-empirical wind-acceleration model in VY\,CMa constrained from multi-epoch observations of H$_2$O masers \citep{richards}. This model assumes linear wind acceleration within 75--440\,mas from the star and does not take into account the inhomogenities of the wind at these scales. Ignoring the weak pedestal, the main TiO component has a full width of about 40$\pm$5\,\kms, implying that the emission region extends out to $r$=28$\pm$10\,R$_{\star}$ (R$_{\star}$=2000\,R$_{\sun}$) or 0\farcs18$\pm$0\farcs07 at 1.2\,kpc and the emission corresponding to the broad pedestal is located at even larger distances from the star. 


With the average deconvolved size of the of TiO$_2$ emission of FWHM=0\farcs6, more than 99\% of the flux is enclosed within a region of the radius of  $r$=3$\sigma/2$=$3 \sqrt{2 \ln{2}}$\,FWHM=0\farcs4 or $r$=50($\pm 10$)\,R$_{\star}$. Using this value and the size of $r$=28\,R$_{\star}$ for TiO, we can correct the column densities derived in Sect.\,\ref{rot-diagrams} for the corresponding beam filling factors. In this way, we obtain $N$(TiO)=(6.7$\pm$0.8)$\cdot$10$^{15}$\,cm$^{-1}$ and $N$(TiO$_2$)= (1.8$\pm$0.2)$\cdot$10$^{15}$\,cm$^{-1}$. For a mass-loss rate of 10$^{-4}$\,M$_{\sun}$\,yr$^{-1}$ \citep{danchi} and an average expansion velocity of 10\,\kms\ for TiO and 20\,\kms\ for TiO$_2$, the molecular abundances with respect to the column of hydrogen (assumed to be in molecular form only) are [$f_1$]=$\log (N_{\rm TiO}/N_{\rm H})$=$-8.6$ and [$f_2$]=$\log (N_{{\rm TiO}_2}/N_{\rm H})$=$-9.5$, but these estimates are very approximate.

\subsection{Ti oxides and the formation of inorganic dust}\label{dis2}

Our interferometric observations of TiO and TiO$_2$ allow us to critically examine present ideas about the onset of nucleation 
and the formation of inorganic dust in oxygen-rich objects with high mass-loss rates. In the chemical equilibrium model of \cite{inorganicdust}, TiO exists in the photospheres of cool stars because the bond dissociation energy is high. As a result it is observed in regions with effective temperatures $T_{\rm eff}\!\lesssim$4000\,K as well as in the cooler material just above the photosphere. TiO$_2$ is predicted to form in the reaction of TiO with H$_2$O which is abundant in these environments \citep{royer10}. Owing to the lower bond dissociation energy of TiO$_2$, it is predicted to be more abundant than TiO only at lower temperatures. At temperatures above about 1400\,K, TiO is the dominant carrier of gas phase titanium; below this temperature 
TiO$_2$ is the most abundant carrier of titanium; and at 1100\,K only about 10\% of all titanium-bearing molecules are TiO. At slightly lower temperatures, TiO$_2$  nucleates to produce larger oxides and their clusters thereby initiating the formation of dust ``seeds''.  

Dust formation in VY\,CMa occurs at a radius\footnote{The dust formation radius is poorly defined owing to the asymmetry and probable variability of the star. Its location derived from observations depends on the stellar parameters and dust type that were assumed, explaining in part the broad range of values cited in the literature.} of  0\farcs05--0\farcs12, or 6--15\,R$_{\star}$ 
\citep{monnier,leSidaner96}. The prediction of \citet{inorganicdust} that TiO should be located primarily inside the dust formation zone is inconsistent with our estimate that the size of the TiO emission region of 28\,R$_{\star}$. Unfortunately, our estimate of the rotational temperature ($T_{\rm rot}$=1010$\pm$870\,K, Sect.\,\ref{ana1}) is too uncertain to test the prediction of Gail \& Sedlmayr that $T$\,$\ga$1100\,K, but the relatively large observed line widths of TiO are inconsistent with the model. Although it is plausible that the outflow traced by the TiO gas is already accelerated to large velocities below the dust formation region\footnote{Outflows in luminous cool stars may be driven by radiation pressure absorbed in molecular bands as proposed by \citet[][and references therein]{jorgensen}, or the mass loss might be driven by the convective/magnetic activity of the giant star \citep[e.g.,][]{smith01} which may lead to considerable turbulent motions very close to the star.}, it is more likely that the TiO emission does not originate from inside the dust formation zone, but rather  from a much more extended region. Similarly, the model of \cite{inorganicdust} is in poor agreement with the observed properties of TiO$_2$. The temperature derived from the observations is about 250\,K, but is predicted to be about 4--5 times higher. The TiO$_2$ emission comes from a region ({\it r}$\approx$0\farcs4), that is a few times larger than the size predicted for the dust formation region, although the uncertainties in both are quite large. Moreover, the offset of the peak emission for both molecules with respect to the center of continuum source cannot be easily reconciled with the predictions. 

Inconsistencies are also encountered when the observations are referred to the equilibrium-chemistry model of \citet{SaH}. In their model which includes condensation, emission of TiO or TiO$_2$ only arises from the warm gas ($T \gtrsim$1400\,K) located in a thin layer above the photosphere. Simple equilibrium-chemistry models evidently cannot explain the observations of VY\,CMa. Indeed, molecular abundances calculated for molecular equilibrium have been demonstrated to fail observational tests in AGB stars \citep{cherchneff} and for some species in VY\,CMa \citep{tanen_alo,kami_alo}. A non-equilibrium approach is needed to explain the circumstellar chemistry of supergiants but such models do not exist.    

Absorption bands of TiO are common features in optical spectra of cool photospheres.  Although emission bands of TiO are rarely observed, they are present in VY\,CMa, and very rich TiO emission is seen in the spectra of  U\,Equ and V4332\,Sgr \citep{uequ,kami_v4332} --- two peculiar stars thought to have recently experienced a major cataclysm \citep[a planet-engulfment and a stellar merger, respectively;][]{geballe,tylenda}. Emission of TiO was also identified in several B[e] supergiants \citep{TiOInBe} and a few protostars \citep{hillenbrand}, including some that are known to be explosive objects. Recently, an emission feature of TiO was discovered in the IR (around 10\,$\mu$m) in four S-type Mira variables which are large-amplitude pulsators \citep{TiOinSstars}. A common property of these objects (including VY\,CMa) is the relatively violent environment in which the TiO emission is observed. In VY\,CMa, it has been suggested that the outflow  was formed by episodic events of enhanced mass loss or ``localized mass ejections"  \citep{smith01,humphreys07} most likely involving shocks \citep[cf.][]{kami_alo}. If newly formed dust is destroyed close to the star by shocks, it might give rise to molecular emission at temperatures below $\sim$1100\,K. The observed variability of the optical TiO bands also suggests that the emission is linked to the stellar activity.

However the kinematic separation of TiO and TiO$_2$ implied by the line profiles in Fig.\,\ref{Fig-averprofile}  (i.e., TiO emission is
highest at velocities where the TiO$_2$ emission has local minima) would appear to contradict this interpretation -- one might expect that simple destruction of the dust would release TiO and TiO$_2$ at the same location. A plausible explanation which is consistent with the observations is that when violent shocks release TiO, the TiO$_2$ is destroyed but it reforms when water reacts with TiO farther from the star. 

Another possible explanation for the Ti oxides we observe is incomplete  formation of the dust. The extended reflection nebula confirms that dust has been effectively formed in VY\,CMa for the last 1000\,yr  or more \citep[e.g.,][]{humphreys07}, but owing to the high mass-loss rate ($\sim$10$^{-4}$\,M$_{\sun}$\,yr$^{-1}$) a sufficient fraction of TiO and TiO$_2$ might survive the nucleation process to account for the observed abundances in the cooler region. The apparent stratification of TiO and TiO$_2$ that we observe might then be understood within the context of the model of \cite{inorganicdust}.  In addition, equilibrium-chemistry models of \citet{SaH} show that the titanium oxides remain relatively abundant throughout the outflow if no condensation takes place. VY\,CMa may be an intermediate case between full versus no condensation. 

All of the models referred to here predict that titanium is converted to the oxides in the inner outflow, but emission features of atomic Ti are observed in the visible at the systemic velocity \citep{waller86,waller2001}. By examining a pure emission feature of \ion{Ti}{I} at 5193\,\AA\ in the VLT spectrum, we find that it covers the range from 0 to 45\,\kms, i.e., is slightly broader than the central component of the radio TiO emission (see Fig.\,\ref{Fig-optical-comp}). This implies that gas-phase \ion{Ti}{I} extends farther from the star than the gas with highest TiO abundance.  
It has long been known that atomic Ti is highly depleted in the diffuse interstellar medium \citep[e.g.,][]{Ti_depletion}, implying that Ti either forms molecules and remains in the gas phase, or it is effectively locked into dust. The observed optical emission of Ti is further evidence that the condensation of Ti species in VY\,CMa is not as efficient as previously estimated, or that the Ti species extend to much larger radii than commonly assumed. Any future attempts to explain this anomaly should take into account those optical observations of atomic titanium. At this moment the observational characteristics of Ti-bearing species in VY\,CMa remain unexplained.

\subsection{Prospects}
The detection of TiO and TiO$_2$ in the cooler region of the CSE of an oxygen-rich late-type star, implies that other small transition-metal bearing molecules such as VO, ScO, YO, and CrO might be found with sensitive interferometers in the (sub-)millimeter wave band.  Observations of TiO and TiO$_2$ with the Atacama Large Millimeter Array (ALMA) when the full sensitivity and angular resolution is attained, should resolve the region where they are present and establish whether there is evidence for stratification between species.    
This in turn would allow more detailed studies of titanium depletion to oxides and solids.

\begin{acknowledgements}
HSPM thanks the Bundesministerium f\"ur Bildung und Forschung (BMBF) for support via project FKZ\,50OF0901 (ICC\,HIFI \textit{Herschel}). The Submillimeter Array is a joint project between the Smithsonian Astrophysical Observatory and the Academia Sinica Institute of Astronomy and Astrophysics and is funded by the Smithsonian Institution and the Academia Sinica. IRAM is supported by INSU/CNRS (France), MPG (Germany) and IGN (Spain). We acknowledge these institutions.  
\end{acknowledgements}


\begin{appendix}
\section{The PdBI spectrum of VY\,CMa} \label{ap1}
\begin{figure*}\centering
\includegraphics[angle=270,width=\textwidth]{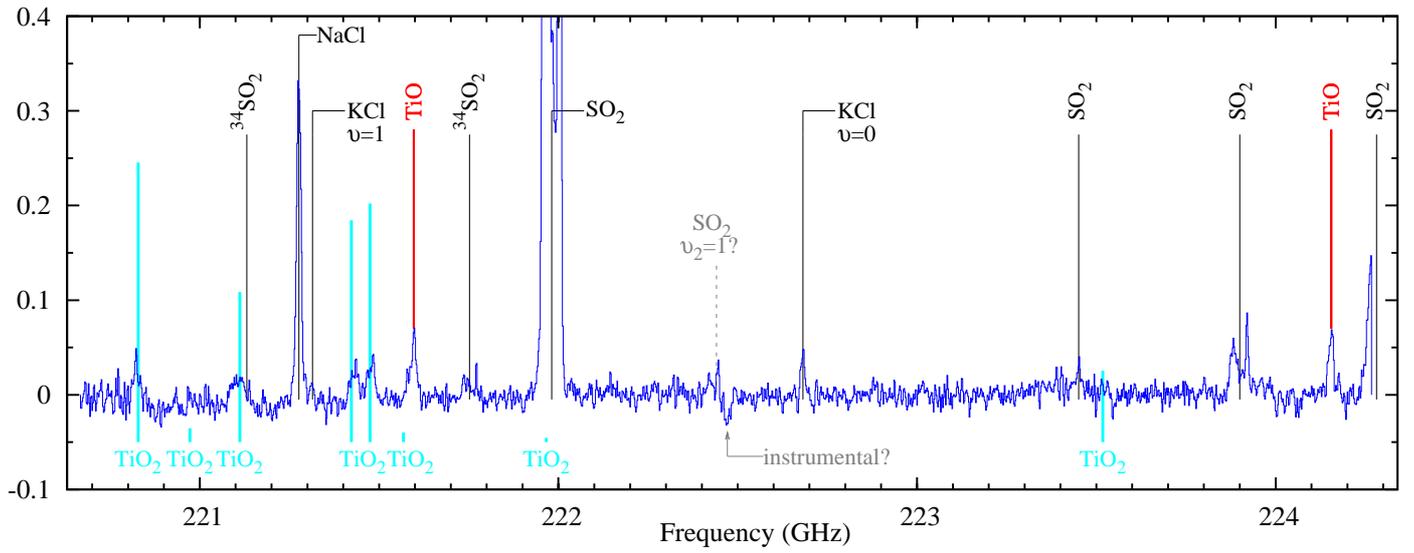}
\caption{The spectrum obtained with PdBI and corresponding to the central pixel of 0\farcs65$\times$0\farcs65. The continuum was subtracted. The height of the lines marking TiO$_2$ positions (cyan), corresponds to the \emph{relative} intensities for LTE at 300\,K. The ordinate is flux density in Jy/beam.}
\label{Fig-PdBI}
\end{figure*}

A spectrum extracted from the central pixel of the PdBI data is displayed in Fig.\,\ref{Fig-PdBI}.\footnote{The spectrum will be also available in electronic form at the CDS via anonymous ftp to
cdsarc.u-strasbg.fr (130.79.128.5) or via http://cdsweb.u-strasbg.fr/cgi-bin/qcat?J/A+A/??????.} The line identification for this spectrum was based mainly on the results of the SMA survey, but also on some archival SMA data \citep[e.g.,][]{fu} and that of \citet{tanen10} were consulted. The identified lines are shown in Fig.\,\ref{Fig-PdBI} and listed in Table\,\ref{Tab-PdBI}. In addition to the targeted TiO and TiO$_2$ lines, the spectrum covers emission features of SO$_2$, $^{34}$SO$_2$, NaCl, and KCl. Among these, KCl had not been reported in VY\,CMa before (a full analysis of its emission will be presented elsewhere). Near 222.45\,GHz, emission and absorption features that seem to compose a P-Cyg profile are seen. The emission peak can be tentatively identified with vibrationally excited SO$_2$ at $\varv_2$=1, but the absorption occurs where two subunits of the WideX correlator are connected and is likely a spurious feature. 

\begin{table} 
\caption{Lines identified in the PdBI spectrum.}\label{Tab-PdBI}
\begin{tabular}{@{}ccc rcc@{}}
\hline
Mole-&
Transition\tablefootmark{a}&
$\nu_{\rm rest}$&
\multicolumn{1}{c}{$E_{u}$}& 
$S_{ul}$\tablefootmark{b}&
$V_{\rm cen}$\tablefootmark{c}\\
cule&&(GHz)&\multicolumn{1}{c}{(K)}&&(\kms)\\           
\hline\hline
      TiO$_2$& 23$_{5,19}\!\to$23$_{4,20}$  & 220.812 & 230 & 12.857 &  14.8\\
      TiO$_2$&  4$_{4,0}\!\to$ 3$_{3,1}$    & 221.096 &  25 &~~3.476 &  17.1\\
$^{34}$SO$_2$& 22$_{2,20}\!\to$22$_{1,21}$  & 221.115 & 248 & 12.663 & --8.9\\
         NaCl&       $J$=17$\to$16    & 221.260 &  96 & 14.200 &  21.7\\
          KCl&$\varv$=1, $J$=29$\to$28& 221.298 & 395 & 16.800 &  18.3\\
      TiO$_2$& 32$_{5,27}\!\to$32$_{4,28}$  & 221.407 & 424 & 19.365 &  35.2\\
      TiO$_2$& 28$_{4,24}\!\to$28$_{3,25}$  & 221.458 & 323 & 14.873 &  26.9\\
          TiO& $J$=7$\to$6, $\Omega$=1& 221.580 &  41 & 13.710 &  20.5\\
$^{34}$SO$_2$& 13$_{2,12}\!\to$13$_{1,13}$ & 221.736 &  93 &~~4.887 &  19.4\\
       SO$_2$& 11$_{1,11}\!\to$10$_{0,10}$ & 221.965 &  60 &~~7.711 &  24.7\\
(SO$_2$)&$\varv_2$=1, 11$_{1,11}\!\to$10$_{0,10}$&222.424&821&~~7.949&  23.0\\
          KCl&     $J$=29$\to$28      & 222.666 & 116 & 25.000 &  22.4\\
       SO$_2$& 27$_{8,20}\!\to$28$_{7,21}$  & 223.435 & 504 &~~3.701 &   9.9\\
       SO$_2$&  6$_{4,2}\!\to$ 7$_{3,5}$    & 223.883 &  59 &~~0.433 &  35.0\\
          TiO& $J$=7$\to$6, $\Omega$=2& 224.139 & 181 & 12.854 &  21.4\\
       SO$_2$& 20$_{2,18}\!\to$19$_{3,17}$  & 224.265 & 208 &~~4.615 & ... \\
\hline\end{tabular}\tablefoot{
\tablefoottext{a}{Quantum numbers characterizing the transition in the standard notation.}
\tablefoottext{b}{The quantum mechanical line strength of the rotational transition.}
\tablefoottext{c}{Centroid position of the emission with respect to the rest frequency.}
}
\end{table}
\end{appendix}

\end{document}